\documentclass[prb,aps,twocolumn,amsmath,amssymb,floatfix,superscriptaddress]{revtex4}

\usepackage[dvips]{graphics}
\usepackage{color}
\usepackage{bbold}
\usepackage{physics}
\usepackage{soul}
\definecolor{dred}{rgb}{0.5,0.2,0}
\usepackage[colorlinks=true, citecolor=blue, urlcolor=blue]{hyperref}
\usepackage{tabularx}
\usepackage{graphicx} 
\date{\today}

\begin{document}

\title{Thermoelectric Enhancement via Electronic and Phononic Channels in Staggered and Non-Staggered Dimerized Quantum Ring}

\author{Ranjini Bhattacharya}

\email{ranjinibhattacharya@gmail.com}

\affiliation{Department of Condensed Matter and Materials Physics, S. N. Bose National Centre for Basic Sciences,
JD-Block, Sector III, Salt Lake, Kolkata 700098, India}

\author{Souvik Roy}

\email{souvikroy138@gmail.com}

\affiliation{School of Physical Sciences, National Institute of Science Education and Research, Jatni 752050, India}
\affiliation{Homi Bhabha National Institute, Training School Complex, Anushaktinagar, Mumbai 400094, India}

\begin{abstract}
Harnessing the quantum coherence and tunability of molecular-scale structures, we theoretically explore thermoelectric transport in ring-shaped molecular junctions featuring dimerized hopping integrals. By engineering alternating strong and weak bonds in both staggered and non-staggered configurations, we reveal a marked transmission asymmetry that drives a substantial enhancement in the thermoelectric figure of merit, $ZT$. To further steer transport behavior, we introduce controlled aperiodicity via site-energy modulations in unit cell format governed by the Aubry-André-Harper (AAH) potential, a quasiperiodic landscape that enables tunable localization-delocalization transitions. This interplay between hopping dimerization and AAH-type disorder gives rise to energy filtering effects and a rich spectrum where extended and critical states coexist, amplifying the Seebeck coefficient while preserving finite electrical conductance. Through a comprehensive non-equilibrium Green’s function analysis, we uncover how key device parameters, including disorder strength, dimerization amplitude, and lead-ring connectivity, collectively shape transport characteristics. Notably, asymmetric lead couplings are shown to enhance performance by leveraging quantum interference pathways. Our findings highlight a robust design strategy for optimizing nanoscale thermoelectric functionality, providing actionable insights for experimental realization in molecular electronic platforms.
\end{abstract}


\maketitle

\section{\label{sec:intro}Introduction}

Rapid depletion of conventional fossil fuel reserves, such as coal, petroleum, and natural gas, has intensified the search for sustainable, environmentally benign methods of energy conversion. In this context, thermoelectric materials~\cite{TE1, TE2, TE3}, which convert heat directly into electrical energy through the Seebeck effect~\cite{seebeck1} (and, conversely, pump heat under an applied current via the Peltier effect~\cite{peltier1}), present an attractive avenue for harvesting waste heat from industrial processes, vehicle exhausts and even low-grade ambient temperature gradients. Although the principle of thermoelectricity dates back to the early nineteenth century, growing concerns over carbon emissions and global warming have renewed interest in both the fundamental physics and practical engineering of thermoelectric devices. Thermoelectric modules, being solid state and maintenance-free, offer long service life and high reliability - ideal for future compact and efficient energy conversion systems.

Despite these advantages, the efficiency of thermoelectric energy conversion is fundamentally constrained by competing requirements: high electrical conductivity, a large Seebeck coefficient, and low thermal conductivity. These metrics are combined in the dimensionless figure of merit.
\begin{equation}
    ZT \;=\; \frac{S^{2}\,G\,T}{K_{\mathrm{el}} + K_{\mathrm{ph}}},
    \label{eq:ZT_intro}
\end{equation}
where $T$ is the absolute temperature, $S$ denotes the Seebeck coefficient, $G$ is the electrical conductance and $K_{\mathrm{el}}$ ($K_{\mathrm{ph}}$) represents the electronic (phononic) contribution  to the total thermal conductance. Achieving a high $ZT$ requires maximizing the power factor $S^{2}G$ while suppressing the parasitic heat flow carried by the phonons. However, in bulk conductors, the Wiedemann-Franz~\cite{wf1, wf2, wf3, wf4} law enforces a proportionality between electronic thermal conductivity and electrical conductivity, making it difficult to enhance $G$ without increasing $K_{\mathrm{el}}$. Consequently, many state‐of‐the‐art bulk thermoelectric materials plateau at $ZT \sim 1$ near room temperature, and although some compounds exceed $ZT=2$~\cite{zt1,zt2,zt3,zt4} at elevated temperatures, scaling this performance to practical, large‐scale devices remains elusive.

Nanostructuring has provided a pathway to partially decouple electrical and thermal transport. In low‐dimensional~\cite{ld1, ld2} systems, such as quantum dots~\cite{qdot}, nanowires~\cite{nw}, superlattices~\cite{sl} and two-dimensional materials, boundary and interface scattering can significantly reduce $K_{\mathrm{ph}}$ without severely degrading electronic mobility. For example, semiconductor superlattices composed of alternating layers exhibit miniband formation for electrons, while phonons are strongly scattered at heterointerfaces~\cite{c6,c7}. One‐dimensional nanowires and carbon nanotubes~\cite{nanotube} leverage boundary scattering to suppress the thermal conductivity of the lattice. Quantum dots, with their discrete energy levels, act as energy filters, selectively transmitting charge carriers near resonant energies and thereby boosting the Seebeck coefficient. In molecular junctions and single‐molecule devices, the energy‐dependent transmission function, $\mathcal{T}(E)$, can be tuned via gate voltages or chemical functionalization, enabling sharp transmission resonances that enhance thermopower~\cite{thermo1, thermo2}. Despite these promising approaches, reproducibly achieving $ZT>2$ in experiment is hindered by factors such as contact resistance~\cite{contres1, contres2}, fabrication variability, and uncontrolled phonon leakage~\cite{phonleak}.

Theoretical model systems have been instrumental in elucidating mechanisms for enhanced thermoelectric performance. A paradigmatic example is the Su–Schrieffer–Heeger (SSH) model\cite{ssh1,ssh2,ssh3,ssh4}, originally proposed to describe soliton excitations in polyacetylene~\cite{ssh1,c4,c5}. The SSH Hamiltonian describes a one‐dimensional tight‐binding chain with alternating (dimerized) nearest‐neighbor hopping amplitudes $t_{1}$ and $t_{2}$. This dimerization gives rise to two topologically distinct phases~\cite{c8,c9}, separated by a band‐inversion transition at $t_{1}=t_{2}$. Although the minimal SSH model neglects electron–electron interactions and disorder, it captures essential physics such as topological edge states~\cite{edge} and fractional charge when the Fermi level lies within the gap. Extensions of the SSH chain that include on‐site potential modulations or longer‐range hopping have revealed connections to phenomena like Thouless charge pumping~\cite{c1} and the mapping between one‐dimensional superlattices and two‐dimensional Hofstadter lattices~\cite{hof}. These studies highlight the interplay between topology~\cite{topo}, band‐structure engineering~\cite{band}, and transport properties in low‐dimensional quantum systems.

Mesoscopic ring geometries~\cite{c2,c3} offer a fertile ground for investigating quantum interference~\cite{qi} effects in charge and heat transport. In such one-dimensional ring systems, the geometry alone can give rise to interference phenomena which may result in pronounced resonances of the transmission function or currents under suitable conditions. The presence of diagonal disorder, whether random or quasiperiodic, typically induces Anderson localization~\cite{Ander}, whereby all eigenstates become exponentially localized in one dimension. Nevertheless, certain types of correlated disorder, such as those described by the Aubry–André–Harper (AAH) potential~\cite{AAH1, AAH2, AAH3, AAH4}, can lead to nontrivial localization transitions and mobility edges, offering an alternative route to study localization phenomena.

\begin{equation}
    \varepsilon_{n} \;=\; \lambda \cos\bigl(2\pi b\,n + \varphi\bigr),
    \label{eq:AAH_intro}
\end{equation}
where $\lambda$ is the modulation amplitude, $b$ is an irrational frequency (commonly chosen as the inverse of the golden ratio)~\cite{gm2}, and $\varphi$ is a phase offset, exhibits a sharp localization‐delocalization transition at $\lambda=2t$ (for nearest‐neighbor hopping amplitude $t$). Below this critical value, eigenstates remain extended; above it, they localize. At the critical point, the spectrum becomes fractal, supporting neither purely localized nor fully extended states. Embedding such an AAH potential into a ring further enriches the localization landscape, enabling controlled crossovers between extended, critical, and localized regimes that strongly influence both electrical conductance and thermopower.

In this work, we analyze the thermoelectric properties of SSH‐type ring structures coupled to external leads in both symmetric and asymmetric configurations. Due to the ring topology, electrons can traverse multiple paths whose relative phases depend on contact positions. By attaching source and drain electrodes to inequivalent lattice sites, we engineer interferometric pathways that produce energy‐dependent transmission asymmetries, $\mathcal{T}(E)\neq \mathcal{T}(-E)$, necessary for obtaining a large Seebeck coefficient. We further superimpose the AAH potential [Eq.~(\ref{eq:AAH_intro})] on each unit cell, allowing us to examine both staggered~\cite{stag1,stag2,stag3} arrangements (where adjacent sites have opposite potential signs) and non‐staggered profiles (where the potential varies smoothly). By varying the modulation amplitude $\lambda$ and the dimerization ratio $t_{1}/t_{2}$, we systematically explore how band‐gap opening (due to hopping alternation) competes with correlated disorder (due to the AAH potential) to influence the coexistence of extended, critical, and localized eigenstates. Because $ZT$ is highly sensitive to sharp features or gaps in the electronic density of states, this interplay can lead to pronounced enhancements in thermoelectric efficiency.

To quantify transport and thermoelectric coefficients, we employ the nonequilibrium Green’s function (NEGF) formalism~\cite{GF1, GF2} in the linear‐response, zero‐bias limit. From the energy‐dependent transmission $\mathcal{T}(E)$, we compute the electrical conductance $G$, the Seebeck coefficient $S$, and the electronic thermal conductance $K_{\mathrm{el}}$. The phononic contribution~\cite{kph1, kph2, kph3, kph4}, $K_{\mathrm{ph}}$, is estimated in the ballistic regime. We investigate both symmetric lead‐ring couplings, where source and drain attach to inversion‐related sites and asymmetric couplings where inversion symmetry is broken to demonstrate how quantum interference can be harnessed to break electron–hole symmetry in $\mathcal{T}(E)$, generating large Seebeck coefficients without severely compromising electrical conductance.

Our study represents the first comprehensive investigation that simultaneously addresses (i) staggered versus non-staggered AAH potentials, (ii) dimerized hopping patterns characteristic of the SSH model, and (iii) geometric asymmetries in lead-ring couplings within a unified thermoelectric framework. We find that specific combinations of AAH modulation strength and dimerization ratio produce sharp transmission antiresonances~\cite{antires} and miniband structures, leading to peaks in the power factor $S^{2}G$ while suppressing total thermal conductance. These features reflect localization‐delocalization crossovers in the eigenstate spectrum, intimately related to the topological properties of the SSH Hamiltonian and the critical behavior of the AAH model. Our results provide clear design guidelines for molecular‐scale thermoelectric devices, showing that engineered bond alternation and correlated disorder in a ring geometry can achieve high $ZT$ values. In particular, asymmetric ring‐lead connections further skew the transmission profile, enabling large Seebeck coefficients without significant loss of electrical conductance.

The remainder of this paper is organized as follows. In Sec.~\ref{sec:model} we introduce the tight‐binding Hamiltonian for the dimerized SSH~ ring with AAH  modulations and outline the NEGF formalism for calculating charge and heat transport. Section~\ref{sec:results} presents a systematic analysis of transport characteristics as a function of dimerization ratio and AAH modulation strength comparing symmetric and asymmetric coupling geometries. We discuss the resulting electrical conductance, Seebeck coefficient, electronic thermal conductance, power factor, and figure of merit $ZT$, highlighting the roles of quantum interference and localization phenomena. Finally, Sec.~\ref{sec:conclusion} summarizes our main findings, elaborates on their implications for molecular‐scale thermoelectric applications, and outlines potential avenues for future research such as incorporating electron–phonon interactions and inelastic scattering effects to approach experimental realism.

\section{Quantum ring, tight binding Hamiltonian and theoretical prescription}

\begin{figure}[ht]
{\centering \resizebox*{7.0cm}{3.5cm}{\includegraphics{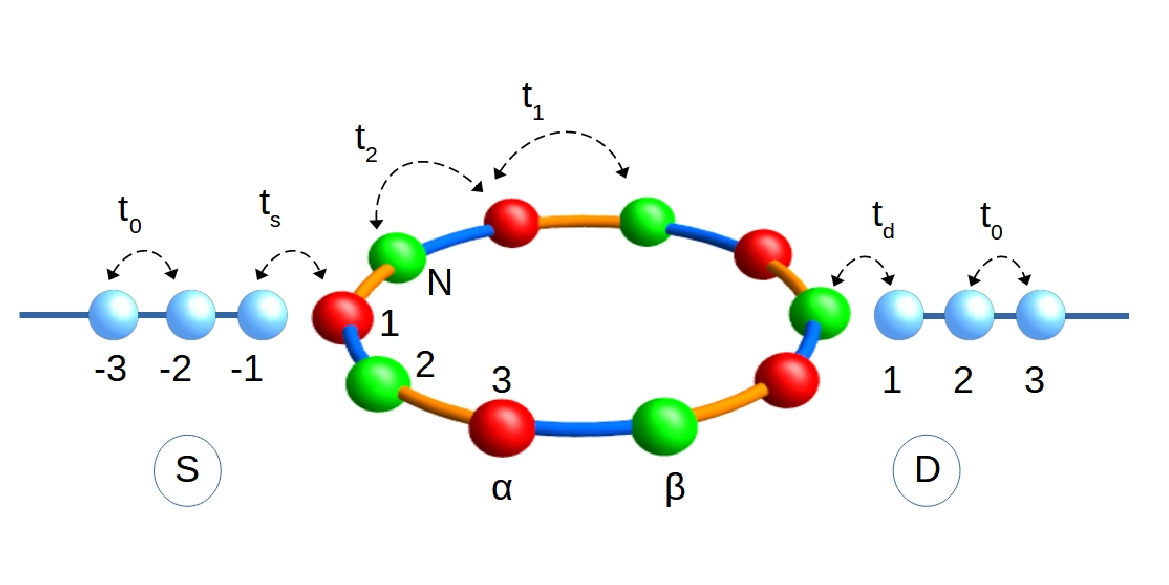}}\par}
\caption{(Color online) The schematic diagram depicts SSH ring symmetrically coupled to source and drain electrodes. In this configuration, the onsite energies are modulated with AAH potential. This setup highlights how the interplay of hopping strengths and AAH potential  provides a framework for effective enhancement of thermoelectric efficiency.}
\label{schematic}
\end{figure}

\section{Model and Theoretical Formalism}
\label{sec:model}

In this section, we present a comprehensive description of the ring geometry employed in our study, along with the formulation of the corresponding tight-binding (TB) Hamiltonian. The model explicitly incorporates a quasiperiodic modulation in the on-site potential, which serves to emulate correlated disorder in a controlled manner. Such a setting allows for the investigation of nontrivial localization phenomena and their influence on quantum transport.

We further outline the methodology adopted to compute the electronic transmission coefficient using the Landauer-Büttiker formalism, which plays a pivotal role in the evaluation of thermoelectric properties. The transmission function, obtained within a fully phase-coherent quantum transport framework, directly enters the expressions for key thermo-electrical quantities such as the electrical conductance, Seebeck coefficient, and electronic contribution to thermal conductance.

Our model is specifically constructed to probe the thermoelectric response of a quantum system subject to correlated disorder, thereby providing insights into the interplay between localization effects and energy-dependent transport. This approach enables a detailed exploration of how the underlying quasiperiodic potential influences the efficiency and performance of thermoelectric conversion at the nanoscale.

\subsection{Quantum Ring Geometry and Tight-Binding Hamiltonian}
\label{subsec:geometry_hamiltonian}

We consider a one-dimensional quantum ring composed of $N$ unit cells arranged periodically to form a closed loop. Each unit cell contains two inequivalent atomic sites, labeled as $\alpha$ and $\beta$, as shown schematically in Fig.~\ref{schematic}. The $\alpha$ and $\beta$ sites are depicted as distinct lattice points to reflect their differing local environments or on-site potentials. This bipartite lattice structure is a hallmark of the Su–Schrieffer–Heeger (SSH) model and introduces an intrinsic hopping asymmetry between sites, enabling the study of topological and localization phenomena in a unified platform.

The system is modeled within a tight-binding framework, where we consider only nearest-neighbor hopping processes. The total Hamiltonian of the isolated ring, in second quantization, is given by:
\begin{align}
H &= \sum_{n=1}^{N} \left( \epsilon_{\alpha,n} \, c_{\alpha,n}^{\dagger} c_{\alpha,n} + \epsilon_{\beta,n} \, c_{\beta,n}^{\dagger} c_{\beta,n} \right) \nonumber \\
&\quad + \sum_{n=1}^{N} \left( t_1 \, c_{\beta,n}^{\dagger} c_{\alpha,n} + t_2 \, c_{\alpha,n+1}^{\dagger} c_{\beta,n} + \mathrm{H.c.} \right),
\label{eq:tb_ring}
\end{align}
where $c_{\alpha,n}^{\dagger}$ ($c_{\alpha,n}$) and $c_{\beta,n}^{\dagger}$ ($c_{\beta,n}$) are the creation (annihilation) operators for electrons on the $\alpha$ and $\beta$ sites of the $n$th unit cell, respectively. The parameters $t_1$ and $t_2$ denote the intra-cell and inter-cell hopping amplitudes, characterizing the dimerized nature of the lattice. The site energies $\epsilon_{\alpha,n}$ and $\epsilon_{\beta,n}$ encode the effect of diagonal modulation, implemented via a deterministic quasiperiodic potential.

To capture the influence of correlated disorder, we incorporate an Aubry–André–Harper (AAH)-type modulation in the on-site energies. Specifically, we define
\begin{align}
\epsilon_{\alpha,n} &= W_{\alpha} \cos(2\pi b n + \phi), \nonumber \\
\epsilon_{\beta,n}  &= W_{\beta} \cos(2\pi b n + \phi),
\label{eq:aah_modulation}
\end{align}
where $W_{\alpha}$ and $W_{\beta}$ are the modulation amplitudes for the $\alpha$ and $\beta$ sites, respectively, $b$ is an irrational number (commonly chosen as $b = \frac{1+\sqrt{5}}{2}$, the inverse of the golden ratio), and $\phi$ is Aubry phase. The irrationality of $b$ ensures quasiperiodicity in the potential, distinguishing it fundamentally from both periodic and random disorder. Unlike uncorrelated randomness, the AAH potential retains long-range spatial correlations, which can give rise to localization transitions at finite modulation strength.

We consider two physically distinct configurations:
\begin{itemize}
    \item \textit{Non-staggered configuration:} $W_{\alpha} = W_{\beta} = W$, where both sublattices experience identical modulations.
    \item \textit{Staggered configuration:} $W_{\alpha} = -W_{\beta} = W$, introducing a relative phase shift between the modulations at the $\alpha$ and $\beta$ sites.
\end{itemize}
The latter case introduces an effective sublattice potential contrast that can significantly alter the spectral and transport characteristics of the system.

\subsection{Lead Coupling and Thermoelectric Setup}
\label{subsec:lead_thermo}

To study thermoelectric transport, the ring is connected to two semi-infinite electrodes acting as thermal reservoirs, referred to as the source and drain maintained at temperatures $T \pm \Delta T/2$, respectively. The electrodes are modeled as one-dimensional tight-binding chains characterized by uniform site energies $\epsilon_0$ and nearest-neighbor hopping integrals $t_0$. These reservoirs are assumed to be reflectionless and ideal, ensuring unidirectional carrier injection and extraction. Coupling between the ring and the leads is restricted to specific lattice sites, typically the first and the $N$th site of the ring to preserve quantum coherence and enforce boundary scattering conditions.

Given that $\Delta T$ is assumed to be small, we operate within the linear-response regime, allowing us to linearize the thermoelectric coefficients around equilibrium. The steady-state charge and heat currents are then computed using the nonequilibrium Green's function (NEGF) formalism or the Landauer–Büttiker~\cite{lb} approach, both of which rely on the energy-dependent transmission function $\mathcal{T}(E)$, which encodes the full quantum interference profile of the system.

The overall transmission and hence the thermopower is highly sensitive to both the symmetry of the coupling geometry and the underlying spectral features induced by the AAH modulation and dimerization. In particular, asymmetric coupling (i.e., attaching leads at inequivalent sublattice sites) can break electron–hole symmetry in $\mathcal{T}(E)$, a necessary condition for nonzero Seebeck coefficients in particle–hole symmetric systems.

The combination of topological dimerization (via $t_1 \neq t_2$), correlated AAH disorder (via Eq.~\eqref{eq:aah_modulation}) collectively determine the thermoelectric performance and coherence-driven properties of the ring system, which we explore in detail in the subsequent sections.

\subsection{Theoretical Framework}

\subsubsection{Transmission Probability via Green’s Function Formalism}

To investigate the thermoelectric characteristics of the system, it is crucial to compute the energy-resolved electronic transmission probability $\mathcal{T}(E)$. In our work, this is evaluated using the non-equilibrium Green’s function (NEGF) approach within the Landauer-Büttiker formalism. The transmission function between two electrodes (source and drain) is given by

\begin{equation}
\mathcal{T}(E) = \operatorname{Tr}[\Gamma_S(E) G^r(E) \Gamma_D(E) G^a(E)],
\label{eq:transmission}
\end{equation}

where $G^r(E)$ and $G^a(E) = [G^r(E)]^\dagger$ denote the retarded and advanced Green’s functions of the central system, respectively. These functions encode the effect of the full system, including coupling to the leads.

The retarded Green’s function of the device region is defined as
\begin{equation}
G^r(E) = \left[E I - H_C - \Sigma_S(E) - \Sigma_D(E)\right]^{-1},
\label{eq:green}
\end{equation}
where $H_C$ is the tight-binding Hamiltonian of the central ring (including the diagonal AAH-type potential), and $\Sigma_{S(D)}(E)$ represent the self-energy matrices of the source (drain) electrodes, encapsulating the influence of the semi-infinite leads on the finite-sized system.

The coupling between the device and the leads is characterized by the broadening matrices
\begin{equation}
\Gamma_{S(D)}(E) = i\left[\Sigma_{S(D)}(E) - \Sigma_{S(D)}^\dagger(E)\right].
\label{eq:broadening}
\end{equation}

These matrices effectively determine the escape rate of electrons from the system to the electrodes and play a pivotal role in determining the line width of resonant states.

\subsubsection{Thermoelectric Response and Figure of Merit}

The performance of a thermoelectric device is evaluated by several key physical quantities: the electrical conductance ($G$), the Seebeck coefficient ($S$), and the electronic contribution to thermal conductance ($K_{\text{el}}$). The linear response regime appropriate when the temperature bias across the device is small. Now all these quantities can be extracted from the transmission function using energy moment integrals of the form

\begin{equation}
\mathcal{L}_n = \frac{2}{h} \int_{-\infty}^{\infty} \mathcal{T}(E) (E - \mu)^n \left(-\frac{\partial f}{\partial E}\right) dE,
\label{eq:landauer-moments}
\end{equation}
where $\mu$ is the chemical potential (typically chosen as the Fermi energy $E_F$), $f(E)$ is the Fermi-Dirac distribution, and $n = 0, 1, 2$ correspond to the different thermoelectric coefficients.

Using the integrals $\mathcal{L}_n$, the transport coefficients can be expressed as:

\begin{align}
G &= e^2 \mathcal{L}_0, \label{eq:conductance}\\
S &= -\frac{1}{eT} \frac{\mathcal{L}_1}{\mathcal{L}_0}, \label{eq:seebeck}\\
K_{\text{el}} &= \frac{1}{T} \left[\mathcal{L}_2 - \frac{\mathcal{L}_1^2}{\mathcal{L}_0} \right], \label{eq:kelectronic}
\end{align}
where $T$ is the average temperature of the system.

In addition to the electronic thermal conductance, the total thermal conductance also includes a phononic contribution $K_{\text{ph}}$, which we compute separately using a lattice dynamics approach and Green's function formalism for phonons. This treatment incorporates the full vibrational spectrum and lattice connectivity of the structure, allowing for an accurate evaluation of phonon-mediated heat transport. Details of this method can be found in Ref.~\cite{phononf}.

Finally, the thermoelectric efficiency of the device is quantified by the dimensionless figure of merit $ZT$, defined as~\cite{li1,li2}

\begin{equation}
ZT = \frac{S^2 G T}{K_{\text{el}} + K_{\text{ph}}}.
\label{eq:zt}
\end{equation}

A high value of $ZT$ (typically $ZT \gtrsim 1$) indicates strong thermoelectric performance. In the quest for efficient nanoscale energy harvesters, achieving such values in quantum-coherent systems with tailored disorder or topological modulations is of significant contemporary interest.

\subsection{Phonon-mediated thermal transport and calculation of $k_{\text{ph}}$}

\begin{figure}[ht]
{\centering \resizebox*{7.0cm}{3.5cm}{\includegraphics{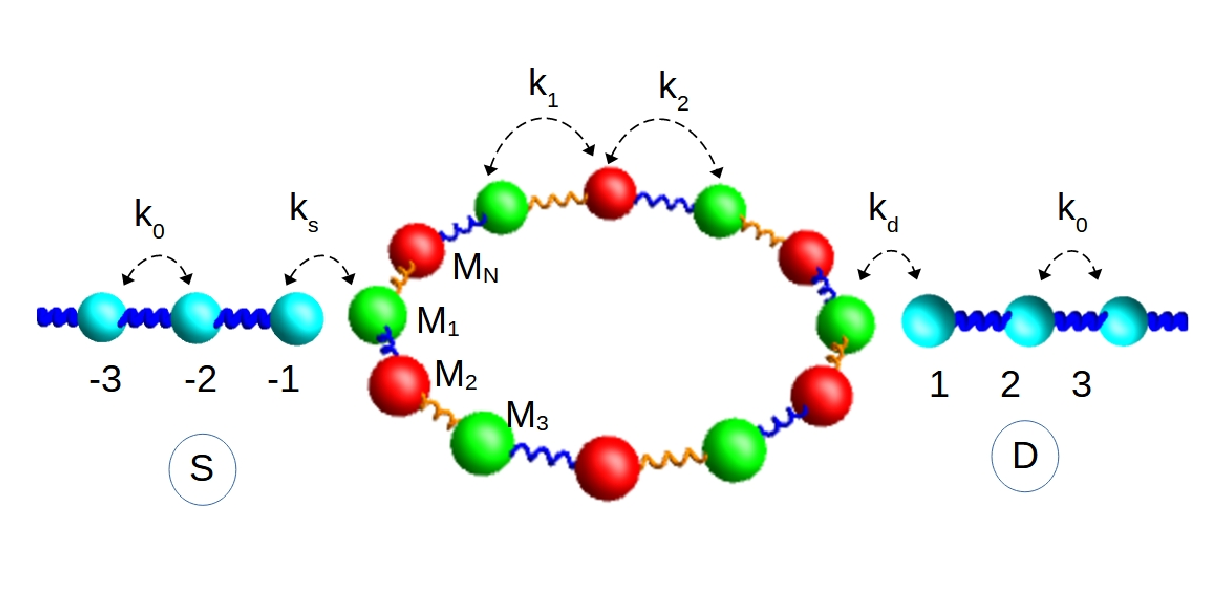}}\par}
\caption{(Color online) The schematic diagram depicts phononic SSH ring symmetrically coupled to source and drain electrodes. In this configuration, the mass is modulated with AAH potential.}
\label{schematic}
\end{figure}

\subsubsection{Theoretical Framework}
Phonons, representing the quantized modes of lattice vibrations, play a crucial role in the transport of thermal energy in crystalline and low-dimensional systems. In the harmonic approximation, a 1D atomic ring can be modeled as a series of point masses connected by springs, wherein each spring-mass unit oscillates about its equilibrium position. The lattice dynamics of such a system are governed by Newton’s second law applied to the $n$-th atom, leading to the following equation of motion:

\begin{equation}
	M_n \omega^2 U_n = V_n U_n - K_{n-1} U_{n-1} - K_{n+1} U_{n+1},
\label{eq2}
\end{equation}

where $M_n$ is the mass of the $n$-th atom, $U_n$ is its displacement from equilibrium, and $K_{n-1}$ and $K_{n+1}$ represent the spring constants of its nearest-neighbor couplings. The on-site restoring term is encapsulated as

\begin{equation}
	V_n = K_{n-1} + K_{n+1},
\label{eq3}
\end{equation}

accounting for the total force experienced by the atom due to its adjacent bonds.

\subsubsection{Green's function framework for evaluating phononic transport}

To evaluate phonon transmission, we employ the non-equilibrium Green’s function (NEGF) approach, a powerful technique rooted in quantum transport theory and particularly suited for mesoscopic systems. The retarded Green’s function $\tilde{G}$ of the central scattering region is given by

\begin{equation}
\tilde{G}(\omega) = \left[\tilde{M} \omega^2 - \tilde{K} - \tilde{\Sigma}_l - \tilde{\Sigma}_r \right]^{-1},
\label{eq6}
\end{equation}

where $\tilde{\Sigma}_{l}$ and $\tilde{\Sigma}_{r}$ are the self-energy matrices that account for the dynamical influence of the left and right phonon reservoirs (leads)~\cite{phononf}. 

The frequency-dependent phonon transmission function $\tau(\omega)$, which encapsulates the probability of phonon modes transmitting through the central region, is then computed as:

\begin{equation}
\tau(\omega) = \mathrm{Tr}\left[\tilde{\Gamma}_r \tilde{G} \tilde{\Gamma}_l \tilde{G}^\dagger \right],
\label{eq9}
\end{equation}

where the coupling (or broadening) matrices $\tilde{\Gamma}_{l,r}$ are defined as

\begin{equation}
\tilde{\Gamma}_{l,r} = i \left( \tilde{\Sigma}_{l,r} - \tilde{\Sigma}_{l,r}^\dagger \right),
\label{eq10}
\end{equation}

representing the energy-level broadening due to reservoir interaction.

\subsubsection{Quantitative assessment of phononic heat transport $k_{\text{ph}}$}

The phonon thermal conductance $k_{\text{ph}}$ at a given temperature $T$ is determined using the Landauer formalism, which relates thermal current to the transmission function. It is given by:

\begin{equation}
K_{\text{ph}} = \frac{\hbar}{2\pi} \int_0^{\omega_c} \tau(\omega) \frac{\partial f_B(\omega, T)}{\partial T} \, \omega \, d\omega,
\label{eq11}
\end{equation}

where $\hbar$ is the reduced Planck constant, $f_B(\omega, T)$ is the Bose-Einstein distribution function, and $\omega_c$ is the maximum vibrational frequency (cut-off frequency) of the system. In the harmonic approximation, $\omega_c$ is estimated as

\begin{equation}
\omega_c = 2 \sqrt{\frac{K}{M}},
\label{eq12}
\end{equation}

reflecting the highest phonon frequency supported by the spring-mass network.

\subsubsection{Implementation of mass modulation via the Aubry-André-Harper potential}

To investigate the influence of quasiperiodicity on phonon transport, we introduce an Aubry-André-Harper (AAH) modulation in the atomic mass distribution. The modulated mass profile is defined as:

\begin{equation}
M_n' = M \left[1 + W \cos(2\pi b n + \phi) \right],
\label{eq13}
\end{equation}

where $W$ characterizes the modulation strength, in similarity with the electronic case $W$ chages as $W_{\alpha} = W_{\beta} = W$ for non-staggared case and as $W_{\alpha} = -W_{\beta} = W$ for staggared case (the red balls denoting the $\alpha$ and green balls denoting the $\beta$ sites respectively) , $b$ is an irrational number (e.g., the inverse golden ratio) to ensure incommensurate structure, and $\phi$ is a tunable phase parameter. This type of modulation induces a deterministic quasidisorder in the lattice, leading to localization effects akin to those observed in disordered systems. Notably, the spring constants $K_n$ and the reservoir self-energies $\Sigma_{l,r}$ are held fixed to isolate the effect of mass modulation.

\section{Numerical Results and Discussion}
\label{sec:results}

In this section, we present a comprehensive numerical investigation of the thermoelectric transport behavior of the proposed model. Our central focus is to identify the parameter regimes that favor enhanced thermoelectric performance, characterized by a high figure of merit ($ZT$). The study systematically explores how different physical parameters such as hopping amplitudes, onsite potential modulation, and structural configurations influence the electronic transport and thermoelectric response.

To maintain consistency and ensure meaningful comparisons, certain key parameters are fixed throughout the simulations unless otherwise stated. Specifically, we set the lead's on-site energies to $\epsilon_i = \epsilon_0 = 0$, with the uniform hopping amplitudes defined as $t_0 = 2$. The coupling strengths between the central scattering region and the source/drain electrodes are taken as $t_S = t_D = 0.8$, ensuring symmetric coupling. The ambient temperature is fixed at $T = 300$~K, corresponding to room temperature conditions. We set the Aubry phase factor $\phi =0$. All energy scales are measured in electron-volts (eV), and any deviation from these standard values is explicitly mentioned where applicable.

The results presented here aim to elucidate the interplay between quantum interference effects, transmission and the resulting thermoelectric efficiency. A special emphasis is placed on identifying how structural modifications such as staggered versus non-staggered modulations and variations in hopping asymmetry affect the transmission function $\mathcal{T}(E)$ and, by extension, the thermoelectric coefficients. This analysis provides fundamental insights into the design of nanostructured systems for optimized energy conversion.

\subsection{Energy-Dependent Transmission Function}

\begin{figure}[ht]
\centering
\resizebox*{8.5cm}{6.5cm}
{\includegraphics{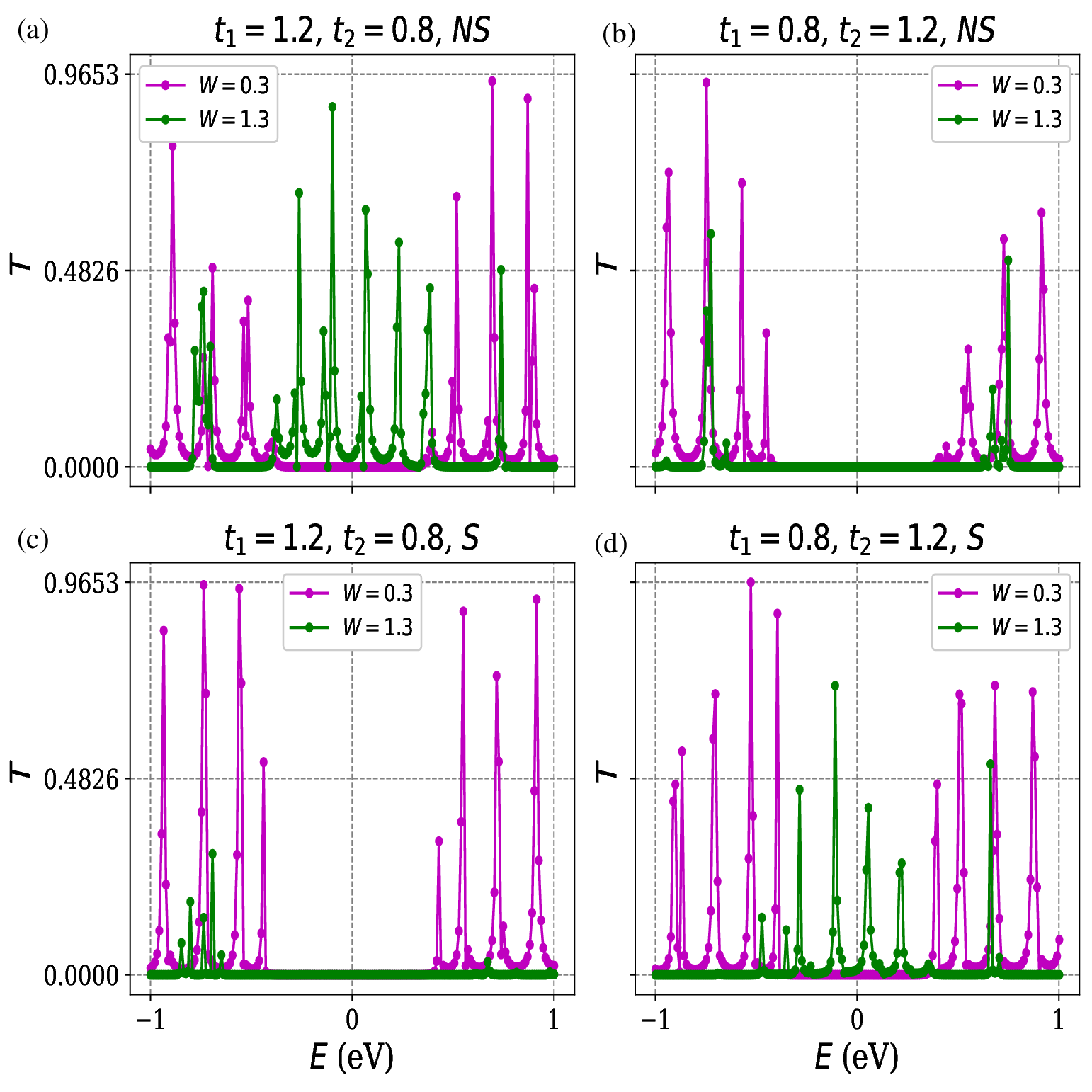}}
\caption{(Color online) Transmission function $\mathcal{T}(E)$ plotted as a function of electron energy $E$. 
(a) and (b) correspond to the non-staggered AAH potential: panel (a) illustrates the case $t_1/t_2 > 1$, while panel (b) shows $t_1/t_2 < 1$, for two onsite potential strengths $W = 0.3$ and $W = 1.3$.
(c) and (d) depict the same scenarios for the staggered AAH potential.
}
\label{fig2}
\end{figure}

We begin our analysis by examining the behavior of the energy-dependent transmission function $\mathcal{T}(E)$ under various structural configurations. Figure~\ref{fig2} provides a comparative illustration of $\mathcal{T}(E)$ across different regimes.

In the top row of Fig.~\ref{fig2}, panels (a) and (b) correspond to a Su-Schrieffer–Heeger (SSH) ring subjected to a non-staggered onsite Aubry-André-Harper (AAH) potential. For panel (a), the hopping asymmetry is chosen such that $t_1/t_2 > 1$. Two cases of potential strength are considered: $W = 0.3$ (pink line) and $W = 1.3$ (green line). It is evident that the transmission profiles differ significantly for these two values. Notably, the transmission spectra span almost the entire energy window, with minimal overlap between the curves, indicating that the electronic transport characteristics are highly sensitive to the onsite potential strength. This broad and energy-spanning transmission is indicative of multiple resonant channels, which is favorable for enhancing the thermoelectric performance.

In contrast, panel (b) depicts the case $t_1/t_2 < 1$ for the same potential strengths. Here, the transmission profiles for $W = 0.3$ and $W = 1.3$ largely overlap, and the transmission peaks are confined to the band edges. The central part of the spectrum exhibits a notable suppression of transmission. This indicates that for the non-staggered case, the configuration with $t_1/t_2 > 1$ offers a more favorable energy window for thermoelectric transport, as it supports stronger and more widely distributed transmission resonances, which directly influence the electrical and thermal conductances, as well as the Seebeck coefficient.

The bottom row of Fig.~\ref{fig2} shows results for the SSH ring when the onsite AAH potential is applied in a staggered manner. In panel (c), for $t_1/t_2 > 1$, the transmission spectra show significant suppression over the energy window, especially as $W$ increases. Conversely, panel (d), corresponding to $t_1/t_2 < 1$, exhibits broadened and more energetically distributed transmission peaks. This is in stark contrast to the non-staggered case and highlights the sensitivity of the system to the modulation pattern of the potential.

From these observations, a noteworthy reversal in optimal transport behavior emerges: while $t_1/t_2 > 1$ favors thermoelectric performance in the non-staggered case, the staggered modulation finds improved transmission characteristics for $t_1/t_2 < 1$. This switching behavior underscores the role of structural symmetry and sublattice modulation in shaping quantum transport pathways.

These findings suggest that tuning the ratio $t_1/t_2$ in conjunction with the nature of the AAH potential (staggered vs non-staggered) provides a powerful knob for engineering transmission asymmetry an essential ingredient for achieving high Seebeck coefficients and, by extension, a large $ZT$. The subsequent sections will further analyze how these transmission features manifest in the thermoelectric performance metrics.

\subsection{Energy-resolved behavior of electronic conductance near the Fermi level}

\begin{figure}[ht]
\centering
\resizebox*{8.5cm}{6.5cm}{\includegraphics{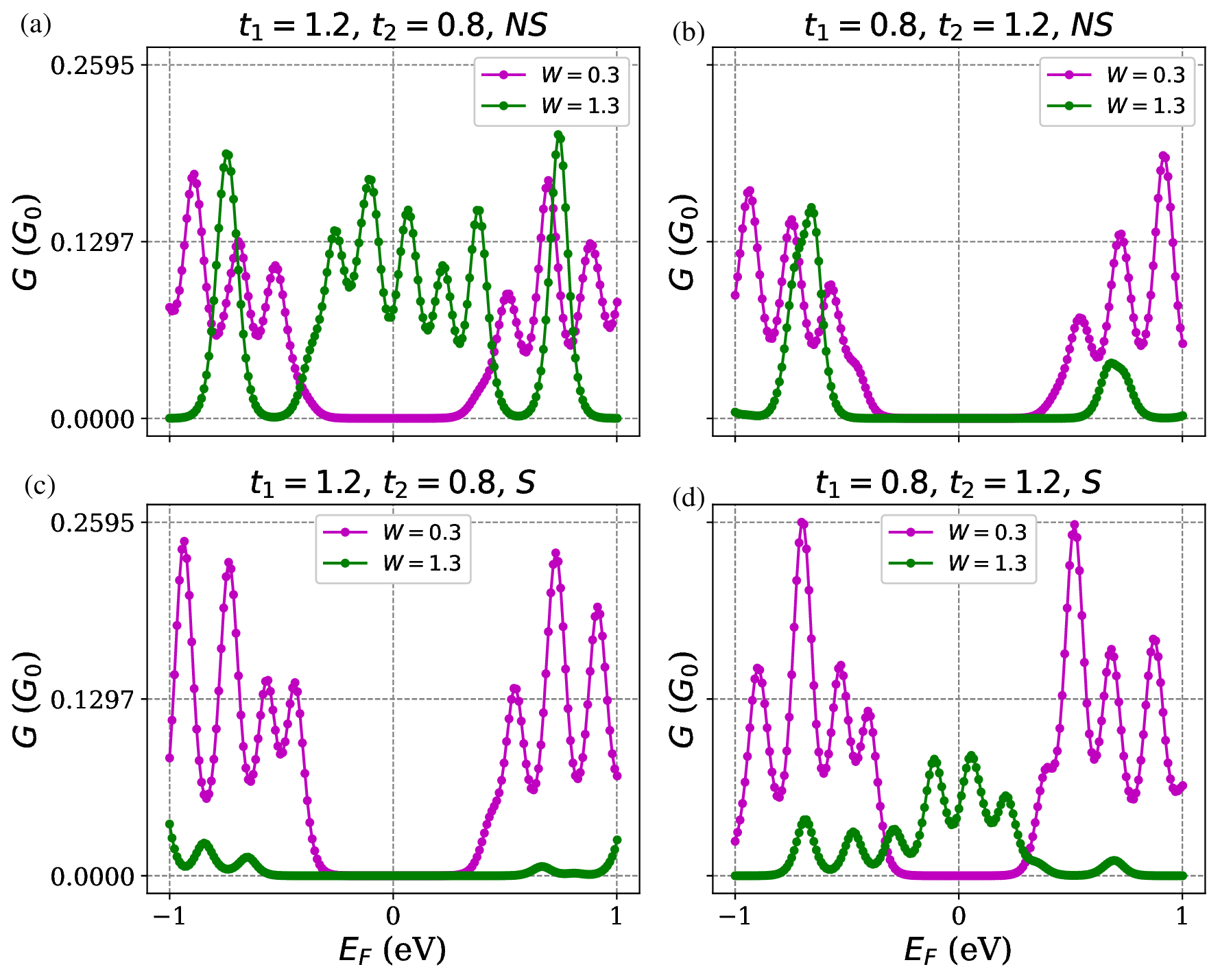}}
\caption{(Color online) Electronic conductance $G$ as a function of the Fermi energy $E_F$. Panels (a) and (b) correspond to non-staggered AAH potentials with $t_1/t_2 > 1$ and $t_1/t_2 < 1$, respectively. Panels (c) and (d) show the corresponding results for the staggered case. In each panel, conductance is plotted for two values of onsite disorder strength: $W = 0.3$ (pink) and $W = 1.3$ (green).}
\label{fig3}
\end{figure}

Figure~\ref{fig3} illustrates the dependence of the electronic conductance $G$ on the Fermi energy $E_F$ for different system configurations. In panels (a) and (b), we consider the non-staggered version of the SSH ring with a modulated Aubry-André-Harper (AAH) potential. Panel (a), corresponding to $t_1/t_2 > 1$, shows that the conductance maintains a relatively high value across a wide range of $E_F$ even at higher disorder strength ($W = 1.3$), comparable to the case with lower disorder ($W = 0.3$). This robust conductance profile suggests that the correlated disorder introduced by the AAH potential, when combined with the topologically nontrivial SSH geometry, supports delocalized electronic states even at high disorder indicating a reentrant delocalization behavior. Such behavior is promising for enhancing thermoelectric performance in nanoscale devices.

In contrast, panel (b) displays the case $t_1/t_2 < 1$, where the conductance profiles for both disorder strengths overlap significantly and are predominantly confined to the band edges. This behavior signifies enhanced localization, consistent with the reduction of topological protection in this regime. 

The lower panels (c) and (d) show the corresponding conductance results for staggered onsite potentials. The trend in panel (c) for $t_1/t_2 > 1$ mirrors the behavior observed in panel (b), while panel (d), representing $t_1/t_2 < 1$, displays features similar to panel (a). These results reflect a duality in conductance response under the exchange of hopping asymmetry and onsite potential arrangement. Despite variations in localization and profile shape, the peak values of $G$ remain nearly constant across all configurations, with a maximum conductance around $0.25G_0$.

\subsection{Fermi energy dependence of the Seebeck coefficient}

\begin{figure}[ht]
\centering
\resizebox*{8.5cm}{6.5cm}{\includegraphics{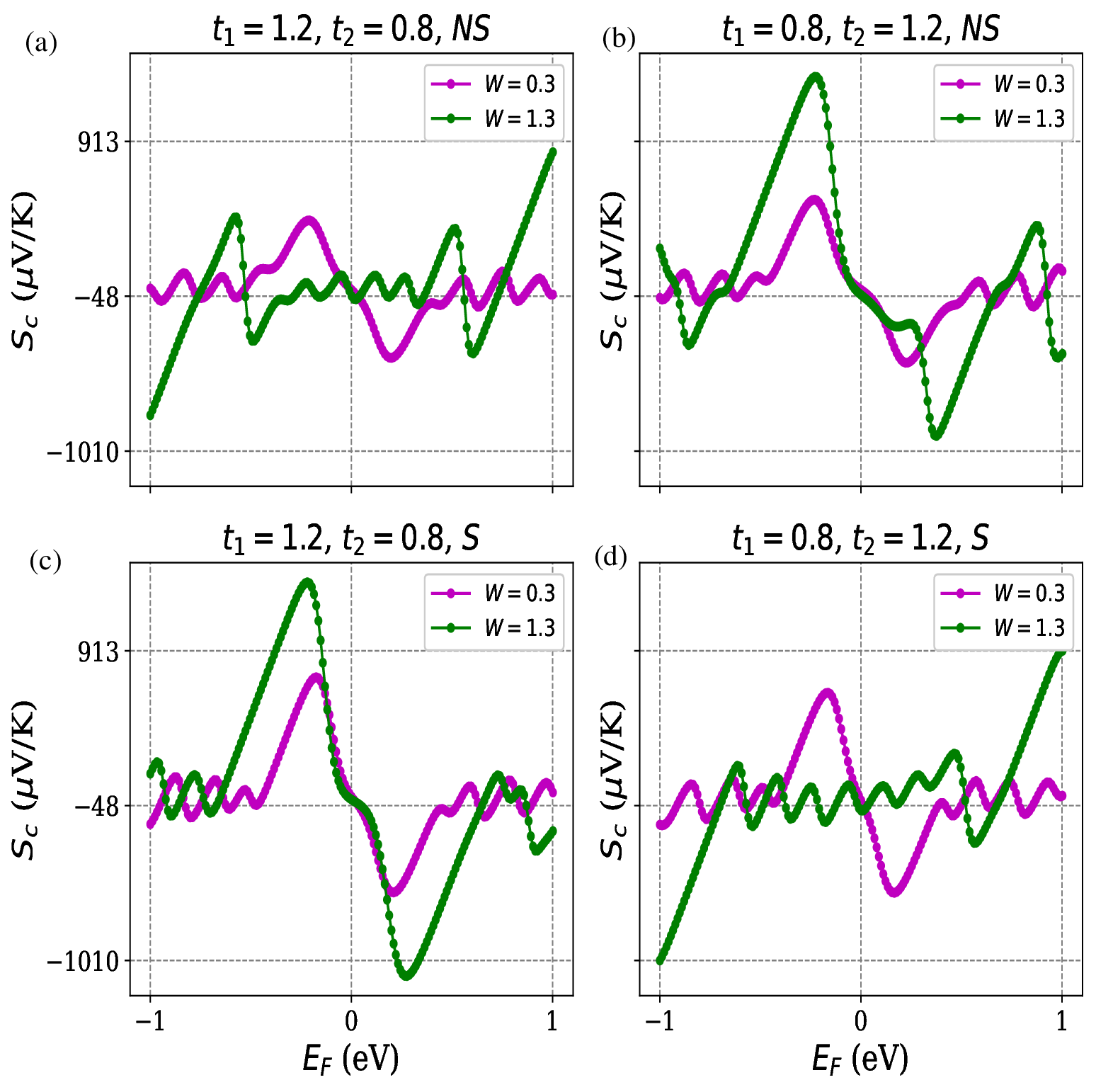}}
\caption{(Color online) Seebeck coefficient $S$ as a function of the Fermi energy $E_F$. Panels (a) and (b) correspond to non-staggered cases with $t_1/t_2 > 1$ and $t_1/t_2 < 1$, respectively. Panels (c) and (d) depict the staggered configurations for the same hopping ratios. The onsite disorder strengths are $W = 0.3$ (pink) and $W = 1.3$ (green).}
\label{fig4}
\end{figure}

In Fig.~\ref{fig4}, we examine the variation of the Seebeck coefficient $S$ as a function of Fermi energy $E_F$ across different model configurations. In the non-staggered scenario shown in panel (a) for $t_1/t_2 > 1$, $S$ exhibits notable peaks in energy regions where the transmission function is suppressed, as previously observed in Fig.~\ref{fig2}(a). This inverse correspondence arises from the dependence of $S$ on the energy derivative (slope) of the transmission function, as expressed in Eqs.~\ref{eq:landauer-moments} and \ref{eq:seebeck}. Specifically, sharp gradients or discontinuities in the transmission lead to significant enhancements in the Seebeck coefficient even in energy windows devoid of high transmission. Consequently, the $S$ profile becomes asymmetric and rich in structure, particularly for the higher disorder case ($W = 1.3$), suggesting favorable thermoelectric response over a wide energy span.

Panel (b), corresponding to $t_1/t_2 < 1$ in the non-staggered configuration, reveals even larger values of $S$ for both disorder strengths. Here, the coefficient rises sharply near the Fermi energies where the transmission begins to deviate from zero, underscoring the critical role of energy-dependent transmission slopes.

In the staggered cases shown in panels (c) and (d), the profiles for $S$ exhibit complementary behavior to their non-staggered counterparts. Notably, the $t_1/t_2 < 1$ case in panel (d) bears close resemblance to the non-staggered $t_1/t_2 > 1$ case (panel a), and vice versa. The maximum values of $S$ reach approximately $1000~\mu\text{V}/\text{K}$ in panels (b) and (c), while panels (a) and (d) produce slightly lower maxima, around $913~\mu\text{V}/\text{K}$. This high magnitude of $S$ underscores the strong potential of these quasi-periodic systems for thermoelectric applications.

\subsection{Electronic thermal transport characteristics as a function of Fermi energy}

\begin{figure}[ht]
\centering
\resizebox*{8.5cm}{6.5cm}{\includegraphics{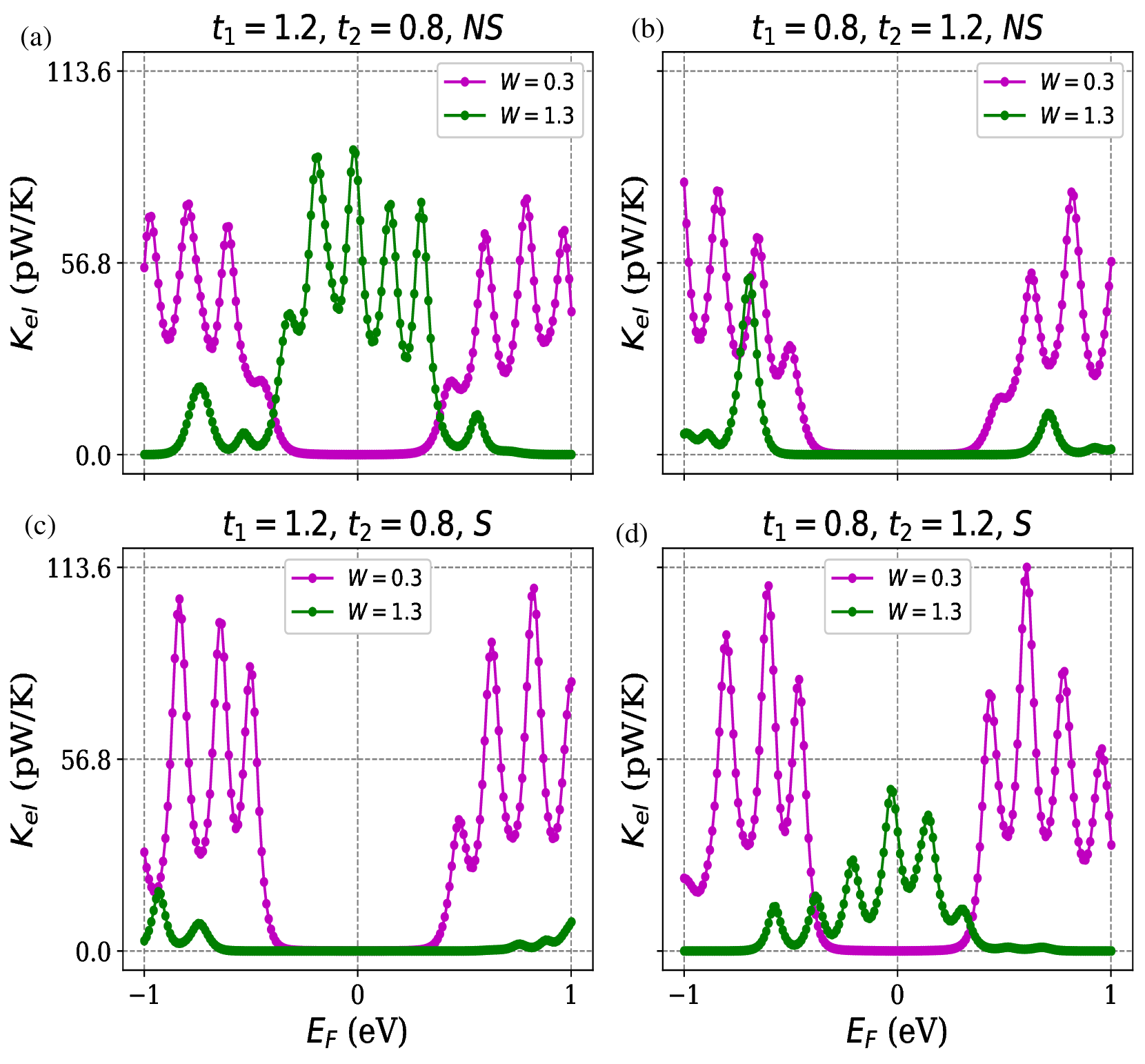}}
\caption{(Color online) Electronic thermal conductance $K_{\text{el}}$ as a function of the Fermi energy $E_F$. Panels (a) and (b) show non-staggered configurations for $t_1/t_2 > 1$ and $t_1/t_2 < 1$, respectively. Panels (c) and (d) depict the staggered configurations. Each panel includes curves for $W = 0.3$ (pink) and $W = 1.3$ (green).}
\label{fig5}
\end{figure}

Figure~\ref{fig5} displays the variation of electronic thermal conductance $K_{\text{el}}$ as a function of the Fermi energy for different disorder and hopping configurations. In panel (a), corresponding to the non-staggered case with $t_1/t_2 > 1$, $K_{\text{el}}$ exhibits distinct behaviors for low and high disorder. While $W = 0.3$ yields a broad distribution of $K_{\text{el}}$ near the band edges, $W = 1.3$ results in mid-band thermal transport. Interestingly, these two curves exhibit limited overlap, effectively covering a wide energy range. The behavior generally parallels the trends in electronic conductance shown in Fig.~\ref{fig3}(a), although deviations particularly for the high-disorder case highlight instances where the Wiedemann-Franz (WF) law does not hold. This breakdown is not uncommon in mesoscopic systems and is considered a key mechanism for enhancing the thermoelectric figure of merit $ZT$.

Panel (b) presents the $t_1/t_2 < 1$ case in the non-staggered configuration. Here, $K_{\text{el}}$ curves for both disorder strengths are largely confined to the band edges and significantly overlap mirroring the localized nature of the eigenstates.

Panels (c) and (d) explore the staggered potential cases. Interestingly, a reversed trend is observed: panel (c) ($t_1/t_2 > 1$) behaves similarly to panel (b), while panel (d) ($t_1/t_2 < 1$) exhibits characteristics akin to panel (a). This reinforces the idea of a duality in thermoelectric response under the inversion of hopping asymmetry and onsite potential staggering. Across all configurations, the peak thermal conductance remains around $113~\text{pW/K}$, underscoring the consistency of transport energy scales across disorder realizations.

\subsection{Variation of thermoelectric figure of merit $ZT$ with Fermi energy}

\begin{figure}[ht]
\centering
\resizebox*{8.5cm}{6.5cm}{\includegraphics{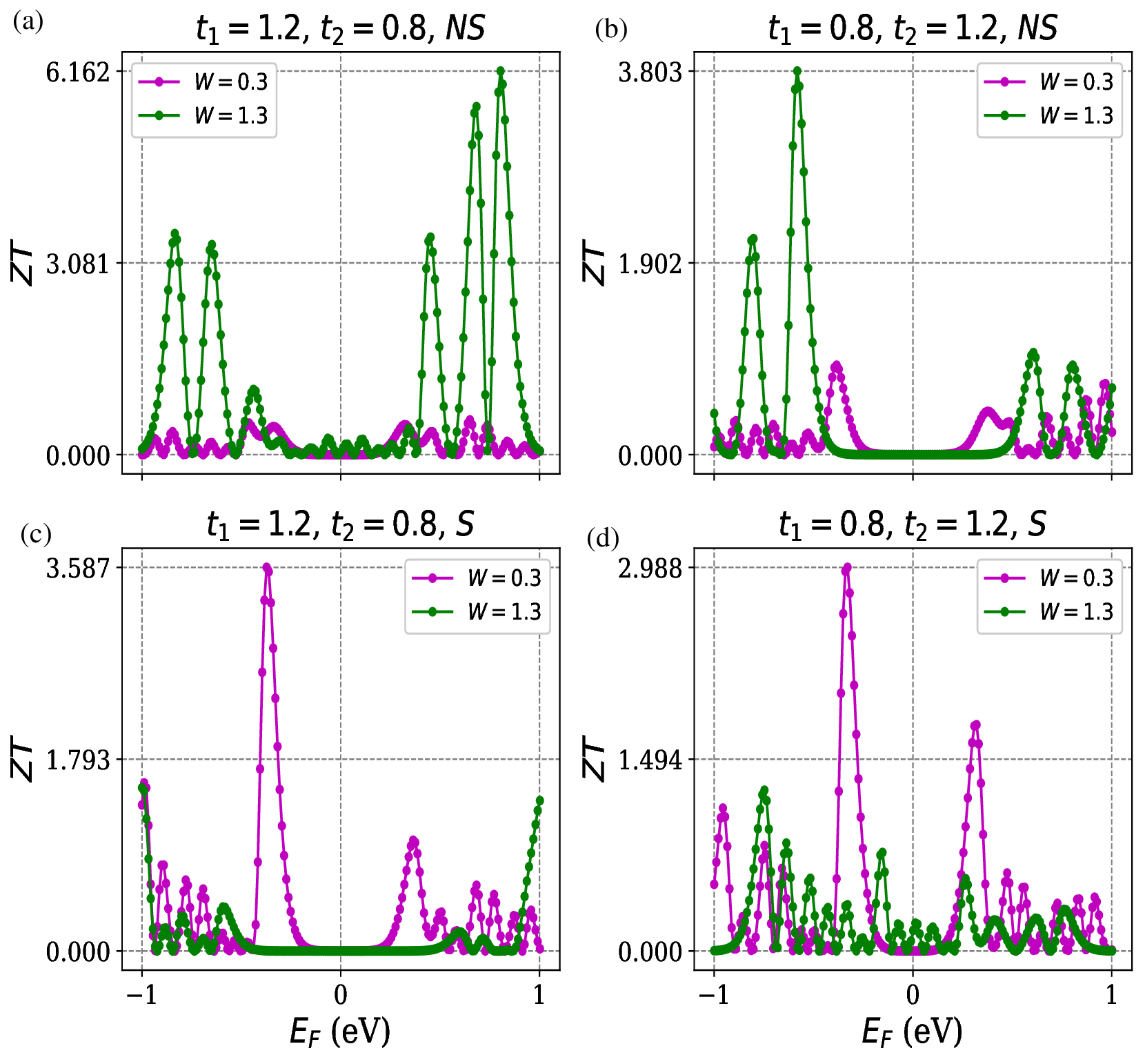}}
\caption{(Color online) Thermoelectric efficiency quantified by the dimensionless figure of merit $ZT$ as a function of the Fermi energy $E_F$. Panels (a) and (b) correspond to the non-staggered configurations with $t_1/t_2 > 1$ and $t_1/t_2 < 1$, respectively. Panels (c) and (d) represent the staggered cases for the same hopping ratios. Each plot includes results for two disorder strengths: $W = 0.3$ (pink) and $W = 1.3$ (green).}
\label{fig6}
\end{figure}

Figure~\ref{fig6} presents the dependence of the thermoelectric figure of merit $ZT$ on the Fermi energy $E_F$, under various hopping and disorder configurations in the Su-Schrieffer-Heeger (SSH) ring modified by a quasiperiodic Aubry-André-Harper (AAH) potential. The top row, panels (a) and (b), illustrates results for non-staggered onsite potentials, whereas the bottom row, panels (c) and (d), shows the staggered counterparts.

In the non-staggered case with $t_1/t_2 > 1$ [Fig.~\ref{fig6}(a)], $ZT$ reaches a remarkably high value of approximately 6 for the stronger disorder strength ($W = 1.3$). This enhancement significantly surpasses the $ZT$ values in the other configurations, where the maximum typically remains around $3$ . Such a high $ZT$ suggests an optimal regime where a combination of low electronic thermal conductance, moderate electrical conductance, and a sharp variation in the transmission function near the Fermi level leads to high Seebeck coefficient and efficient thermoelectric energy conversion.

This regime can be understood as the outcome of a delicate interplay between localization and delocalization of wavefunctions, induced by the non-staggered AAH potential. The aperiodic modulation of onsite energies introduces correlated disorder, which can partially suppress backscattering while maintaining coherent transport through extended states. When this potential is embedded in a topologically nontrivial SSH ring structure with $t_1/t_2 > 1$, the combination fosters mobility edges and energy-dependent localization, thereby optimizing the energy filtering mechanism essential for large Seebeck coefficients and low thermal losses.

In comparison, the other three configurations non-staggered with $t_1/t_2 < 1$ [panel (b)] and both staggered cases [panels (c) and (d)]—display more modest $ZT$ values, peaking around 3 for the same range of disorder. These results affirm that while the staggered arrangements still support finite thermoelectric efficiency, the synergy between the hopping configuration and the nature of the potential plays a decisive role in maximizing $ZT$. The non-staggered configuration with strong dimerization ($t_1/t_2 > 1$) emerges as the most favorable setting for thermoelectric optimization in these quasiperiodic nanostructures.

\subsection{Variation of maximum electrical conductance under increasing on-site disorder}

\begin{figure}[ht]
\centering
\resizebox*{8.5cm}{6.5cm}{\includegraphics{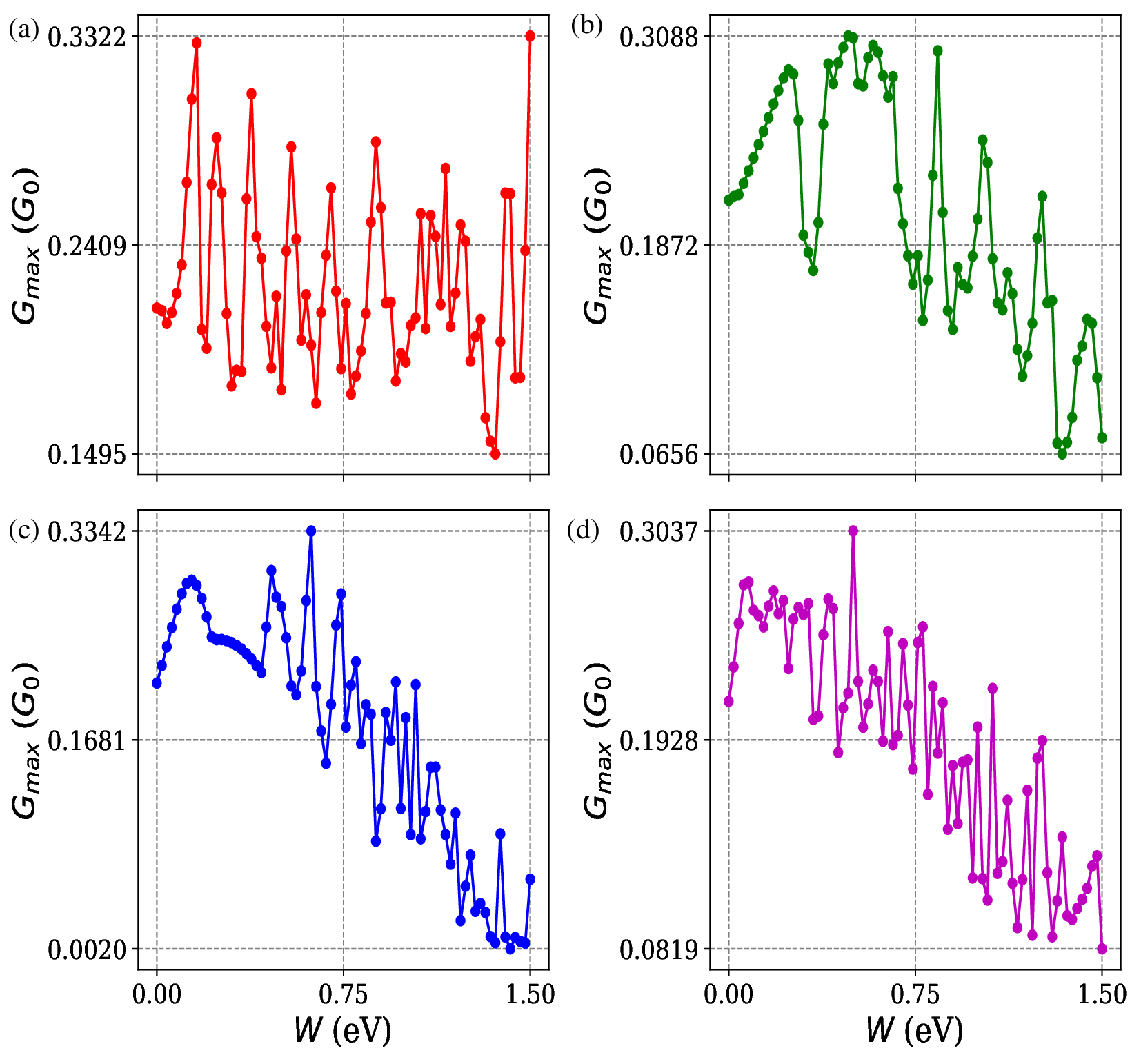}}
\caption{(Color online) Maximum electrical conductance $G_{\text{max}}$ evaluated over the full Fermi energy range, plotted as a function of the onsite disorder strength $W$. Panels (a) and (b) correspond to non-staggered configurations with $t_1/t_2 > 1$ and $t_1/t_2 < 1$, respectively. Panels (c) and (d) show the staggered cases for the same hopping ratios.}
\label{fig7}
\end{figure}

Figure~\ref{fig7} shows how the maximum electrical conductance $G_{\text{max}}$, computed over the entire Fermi energy window, evolves with increasing strength of the quasiperiodic AAH potential. This metric provides a compact yet insightful measure of how efficiently electronic states contribute to transport across a range of energetic configurations.

In the non-staggered setups [panels (a) and (b)], the overall conductance remains much lower, approximately $0.3\,G_0$ and exhibits minimal sensitivity to the hopping asymmetry. This flattening of $G_{\text{max}}$ across the disorder axis implies that in the absence of potential staggering, the system loses its tunability through dimerization, with transport largely dictated by the inherent characteristics of the AAH potential.

In the staggered configurations [panels (c) and (d)], distinct trends emerge depending on the hopping asymmetry. For $t_1/t_2 > 1$ [panel (c)], $G_{\text{max}}$ begins with a relatively high value around $0.33\,G_0$ at weak disorder, but decreases monotonically as $W$ increases. This trend is consistent with the expected suppression of coherent transport in the presence of stronger onsite potential variations. Conversely, the $t_1/t_2 < 1$ case [panel (d)] shows a slightly slower decline in $G_{\text{max}}$, indicating greater resilience to disorder. This suggests that in the staggered configuration, the system with dominant $t_2$ hopping can maintain a broader window of conductive states even under moderate localization effects.

It is particularly noteworthy that the conductance profile in Fig.~\ref{fig7}(a) (non-staggered $t_1/t_2 > 1$) resembles the conductance behavior in the staggered $t_1/t_2 < 1$ case (Fig.~\ref{fig7}(d)). This cross-correlation highlights a form of symmetry or duality in the parameter space, suggesting that equivalent transport features may emerge under reversed hopping and potential configurations. From a design standpoint, this duality enables flexible engineering of thermoelectric responses by toggling the nature of the potential (staggered vs non-staggered) and controlling the dimerization ratio.

These findings underscore the critical role of structural asymmetry and quasiperiodic potential profiles in modulating transport efficiency and, by extension, the overall thermoelectric performance. Careful tuning of disorder and hopping parameters within this minimal SSH-AAH hybrid framework can thus serve as a powerful route toward designing optimized quantum thermoelectric materials.

\subsection{Disorder dependence of maximum Seebeck response in quasiperiodic systems}

\begin{figure}[ht]
\centering
\resizebox*{8.5cm}{6.5cm} 
{\includegraphics{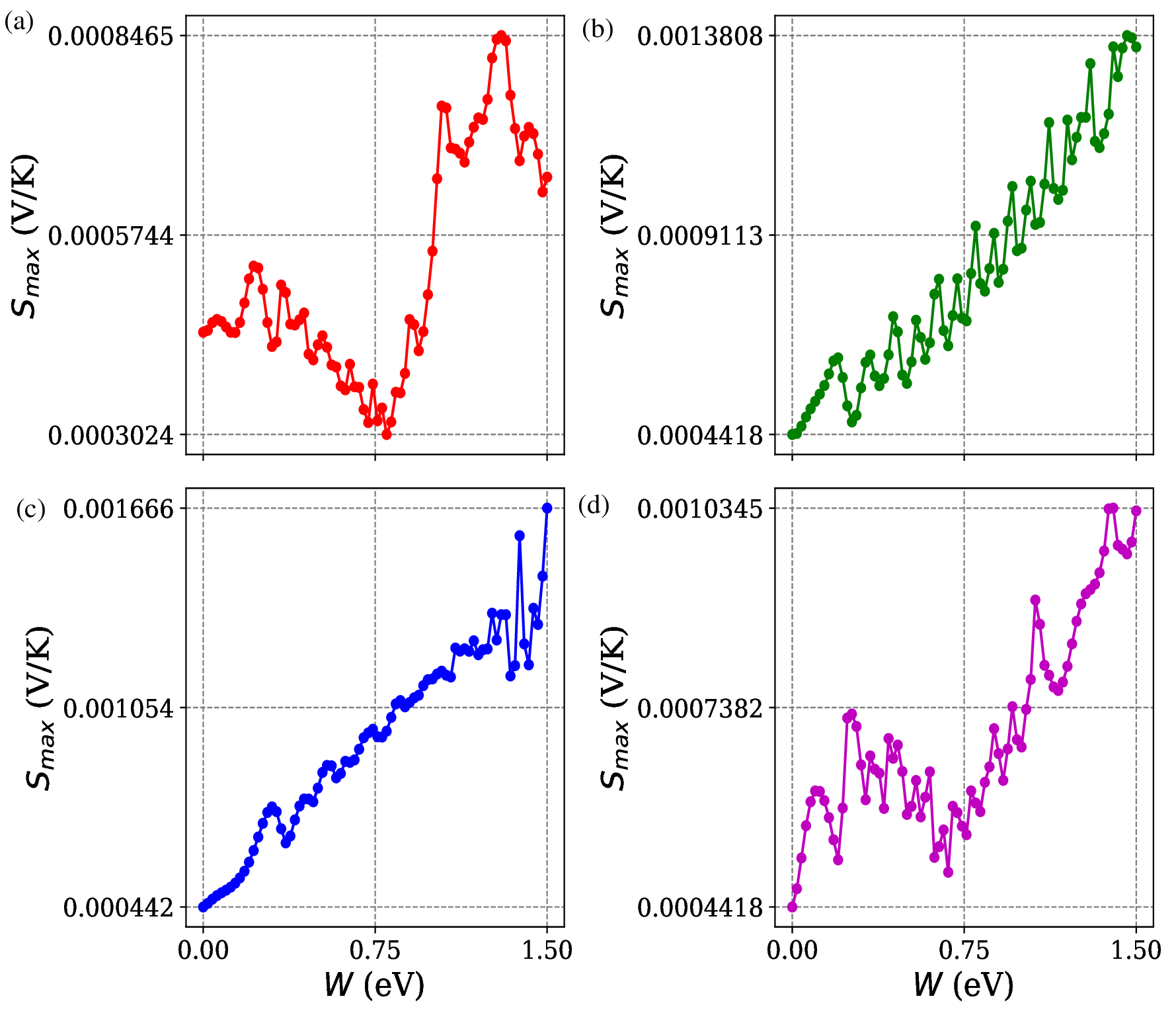}}
\caption{(Color online) Maximum Seebeck coefficient $S_{\text{max}}$ as a function of onsite disorder strength $W$, obtained by scanning over the full Fermi energy window. (a) and (b) correspond to the non-staggered configuration for $t_1/t_2 > 1$ and $t_1/t_2 < 1$, respectively. (c) and (d) show the results for the staggered configuration under the same respective hopping ratios.}
\label{fig8}
\end{figure}

Figure~\ref{fig8} displays the evolution of the peak Seebeck coefficient $S_{\text{max}}$ as a function of the onsite disorder strength $W$, introduced via a correlated Aubry-André-Harper (AAH) potential. The results are presented for both staggered and non-staggered SSH ring configurations under varying hopping asymmetry conditions, $t_1/t_2 > 1$ and $t_1/t_2 < 1$.

In the non-staggered configurations [Figs.~\ref{fig8}(a)-(b)], the qualitative features of $S_{\text{max}}$ as a function of $W$ is shown. Panel (a) depicts $t_1/t_2 > 1$ case where $S_{max}$ shows oscillatory peaks while having decreasing trend with increasing $W$. It gradually reaches to a minimum at $W=0.75$, then rises abruptly and maintaining the oscillatory nature with increasing $W$. This nature of $S_{max}$ hints that with increasing disorder initially the delocalization takes place as a result of which $S_{max}$ falls then the localization increases which enhances the transmission asymmetry and $S_{max}$ rises. In panel (b) $S_{max}$ features different outcome, here $S_{max}$ starts gradually rising with mild oscillations. This suggest the onset of localization with increasing disorder strength. Hence from the non-staggared case we find that how the interplay of hopping terms and disorder strength  impacts on localization-delocalization phenomena and thus control the Seebeck coefficient.       

In the staggered case [Figs.~\ref{fig8}(c)-(d)], where the modulation is applied alternately to the onsite terms, the Seebeck coefficient reaches its maximum in the high-disorder regime ($W \lesssim 1.5$), with values approaching $\approx 1600~\mu\text{V/K}$. This high thermopower arises due to sharp energy-dependent asymmetries in the transmission spectrum, which act as effective energy filters. These asymmetries are often enhanced in quasiperiodic systems due to partial localization and spectral fragmentation. As disorder increases, $S$ exhibits a nonmonotonic trend: it initially decreases near the critical regime ($W \sim 0.75$) in Fig.~\ref{fig8}(d), then shows partial recovery at higher disorder strengths. This reflects a competition between increased localization, which suppresses transport, and enhanced spectral sharpness, which favors high thermopower.

Comparing Figs.~\ref{fig8}(c) and (d), it is observed that the $t_1/t_2 > 1$ configuration yields slightly higher $S_{\text{max}}$ values than $t_1/t_2 < 1$, implying that tuning the hopping asymmetry can be an effective strategy for thermopower enhancement. 

 Notably, a visual correspondence is seen between Fig.~\ref{fig8}(a) and Fig.~\ref{fig8}(d), and likewise between Fig.~\ref{fig8}(b) and Fig.~\ref{fig8}(c), suggesting a form of transport symmetry under inversion of the hopping ratio and staggering condition. This symmetry reinforces the notion that the interplay between lattice symmetry and correlated potential modulations governs the thermoelectric response in SSH-type quasiperiodic rings.

\vspace{0.5cm}

\subsection{Impact of disorder strength on maximum electronic thermal conductance}

\begin{figure}[ht]
\centering
\resizebox*{8.5cm}{6.5cm}
{\includegraphics{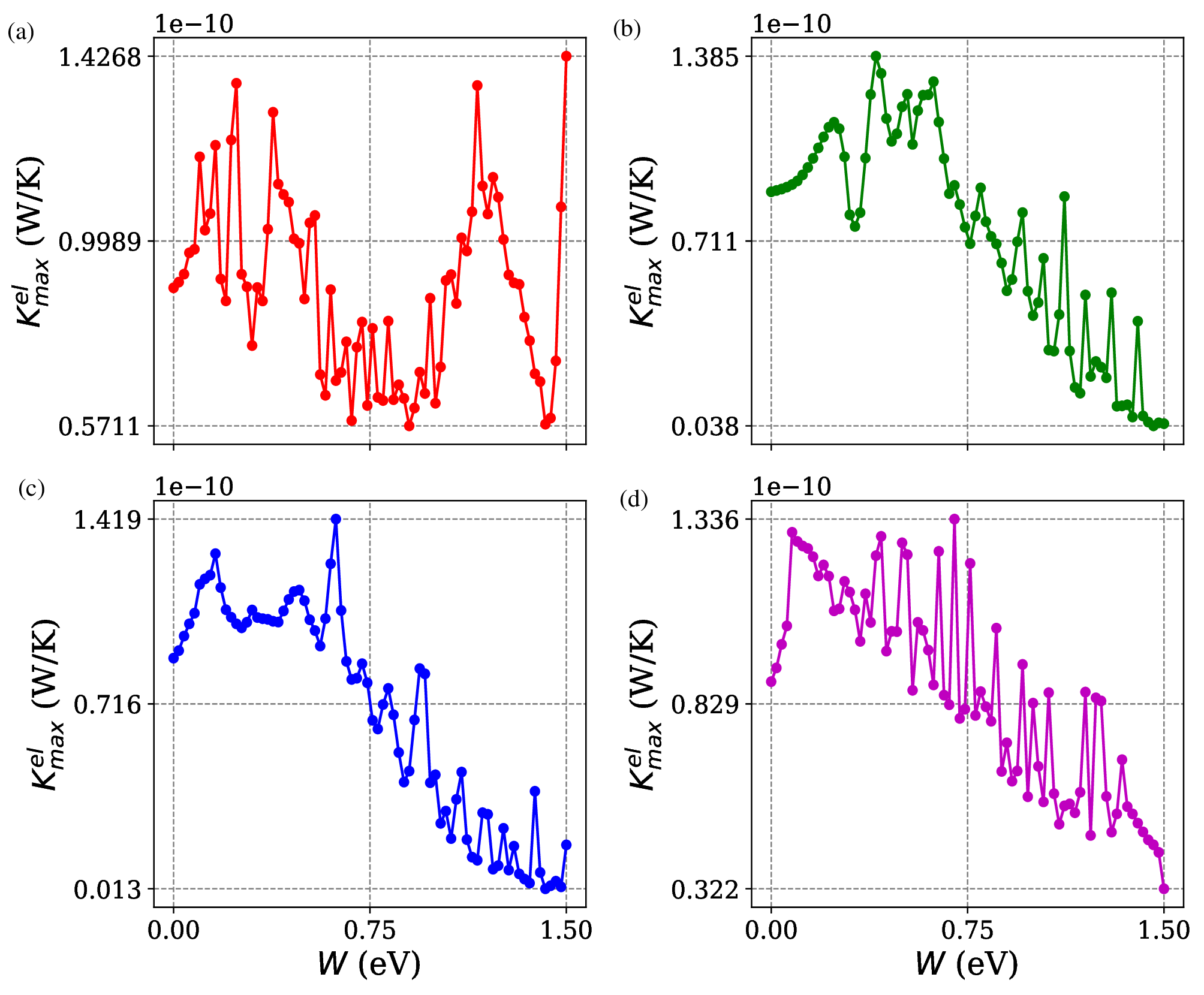}}
\caption{(Color online) Maximum electronic thermal conductance $K^{el}_{\text{max}}$ as a function of disorder strength $W$ for both non-staggered [panels (a)-(b)] and staggered [panels (c)-(d)] configurations, under varying hopping ratios.}
\label{fig9}
\end{figure}

The variation of the electronic thermal conductance $K^{el}_{\text{max}}$ with disorder strength $W$ is depicted in Fig.~\ref{fig9}. The data represent peak conductance values extracted from energy-resolved spectra over the full Fermi window.

In the non-staggered setup [Figs.~\ref{fig9}(a)-(b)], the values of $K^{el}_{\text{max}}$ remain close to $140~\text{pW/K}$ in both hopping configurations. However, the rate of decay with increasing disorder differs—indicating stronger suppression of thermal transport in the $t_1/t_2 > 1$ case. This asymmetry in the localization dynamics across different structural profiles underscores the role of spectral fragmentation and disorder-assisted transport suppression.

In the staggered geometry [Figs.~\ref{fig9}(c)-(d)], we observe distinct trends based on the hopping ratio. For $t_1/t_2 > 1$, $K^{el}_{\text{max}}$ exhibits a nonmonotonic profile, starting from a relatively high value around $143~\text{pW/K}$ at low $W$, then declining to a minimum near $W \approx 0.75$, before showing signs of recovery at higher disorder. This behavior suggests a crossover from a moderately localized to a quasi-extended regime, possibly due to reentrant delocalization effects that can arise in quasiperiodic lattices.

For $t_1/t_2 < 1$, the conductance shows a more conventional decline with increasing $W$, consistent with enhanced localization. The difference in trends between the two cases again reflects how hopping asymmetry influences the energy dispersion and transmission window. A more extended state distribution near the Fermi level can sustain thermal transport better, while localized states contribute minimally.

The intricate variation of $K^{el}_{\text{max}}$ with $W$ directly reflects changes in the shape and width of the transmission function. A sharp, asymmetric transmission window leads to enhanced thermopower, while broad, symmetric transmission contributes more to $K^{el}_{\text{max}}$. Thus, optimizing thermoelectric efficiency demands careful balancing between these opposing tendencies.

\vspace{0.5cm}

\subsection{Disorder-induced modulation of phonon-mediated maximum thermal conductance}

\begin{figure}[ht]
\centering
\resizebox*{8.5cm}{6.5cm}
{\includegraphics{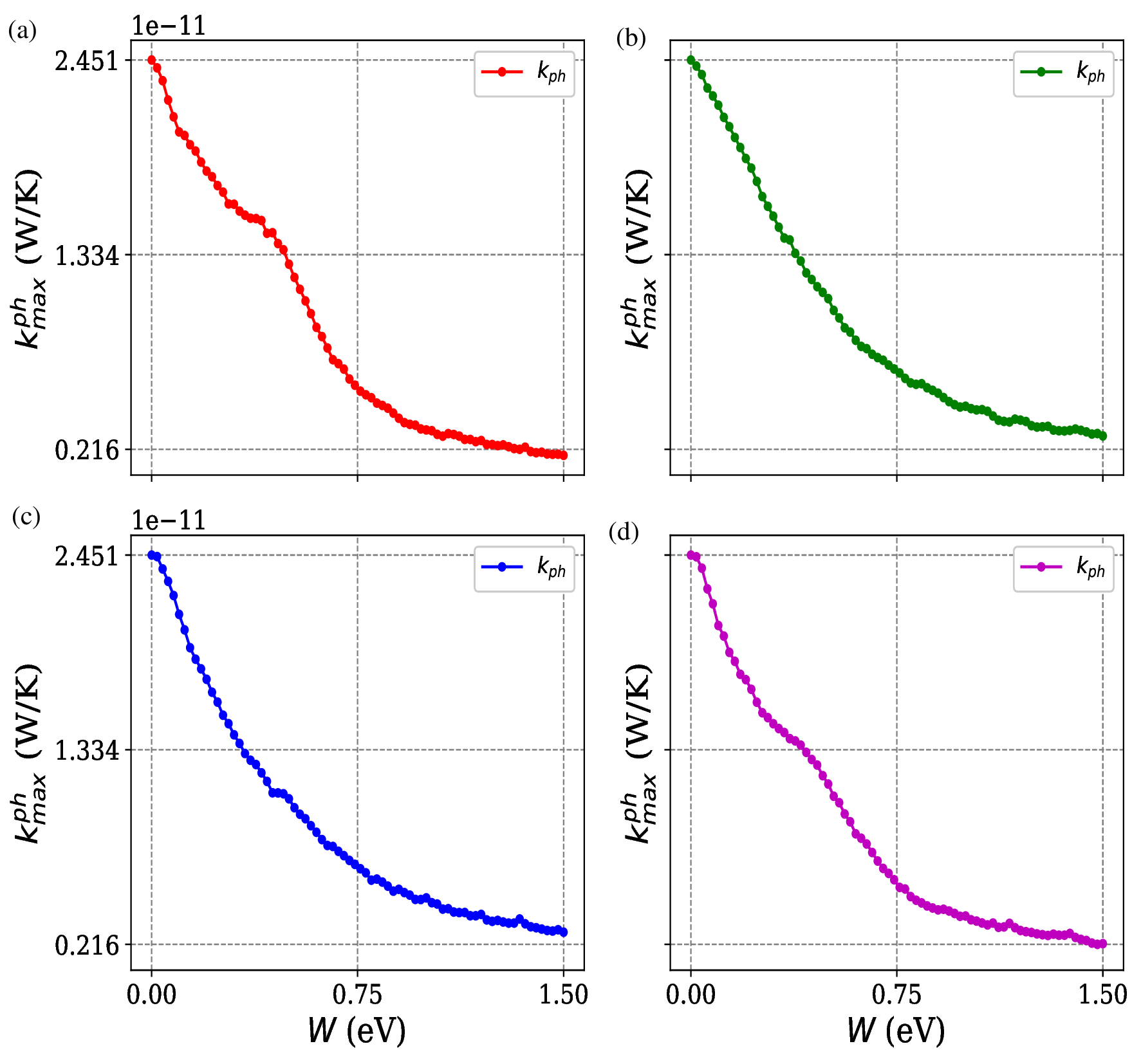}}
\caption{(Color online) Maximum phononic thermal conductance $K^{ph}_{\text{max}}$ as a function of onsite disorder strength $W$ under non-staggered [(a)-(b)] and staggered [(c)-(d)] configurations.}
\label{fig10}
\end{figure}

Phonon-mediated thermal transport is a critical component in determining the thermoelectric figure of merit. In Fig.~\ref{fig10}, we investigate the influence of diagonal correlated disorder on the phononic thermal conductance $K^{ph}_{\text{max}}$, using a spring-mass model mapped onto the SSH ring geometry.

The model considers 50 atoms with mass modulation mimicking the disorder configuration used in the electronic sector. Such correlated mass disorder leads to phonon localization, particularly affecting both low- and high-frequency vibrational modes. This results in strong suppression of phononic heat transport due to increased scattering and mode localization.

Across all configurations staggered and non-staggered, $k_1/k_2 > 1$ and $k_1/k_2 < 1$ the maximum phononic thermal conductance remains around $24~\text{pW/K}$. This indicates that $K_{\text{ph}}$ is less sensitive to hopping asymmetry compared to its electronic counterpart, but remains highly susceptible to disorder-induced scattering.

Interestingly, similar to earlier observations, the $t_1/t_2 > 1$ non-staggered case and the $t_1/t_2 < 1$ staggered case exhibit comparable behavior, again reflecting the transport symmetry present in this class of models. Suppression of $K_{\text{ph}}$ alongside preservation or enhancement of $S$ and $K_{\text{el}}$ is favorable for maximizing $ZT$, highlighting the utility of quasiperiodic SSH rings for phonon-glass electron-crystal-type thermoelectric engineering.

\subsection{Disorder-induced variation of thermoelectric efficiency ($ZT$)}
\label{subsec:zt_disorder_variation}

\begin{figure}[ht]
\centering
\resizebox*{8.5cm}{6.5cm}
{\includegraphics{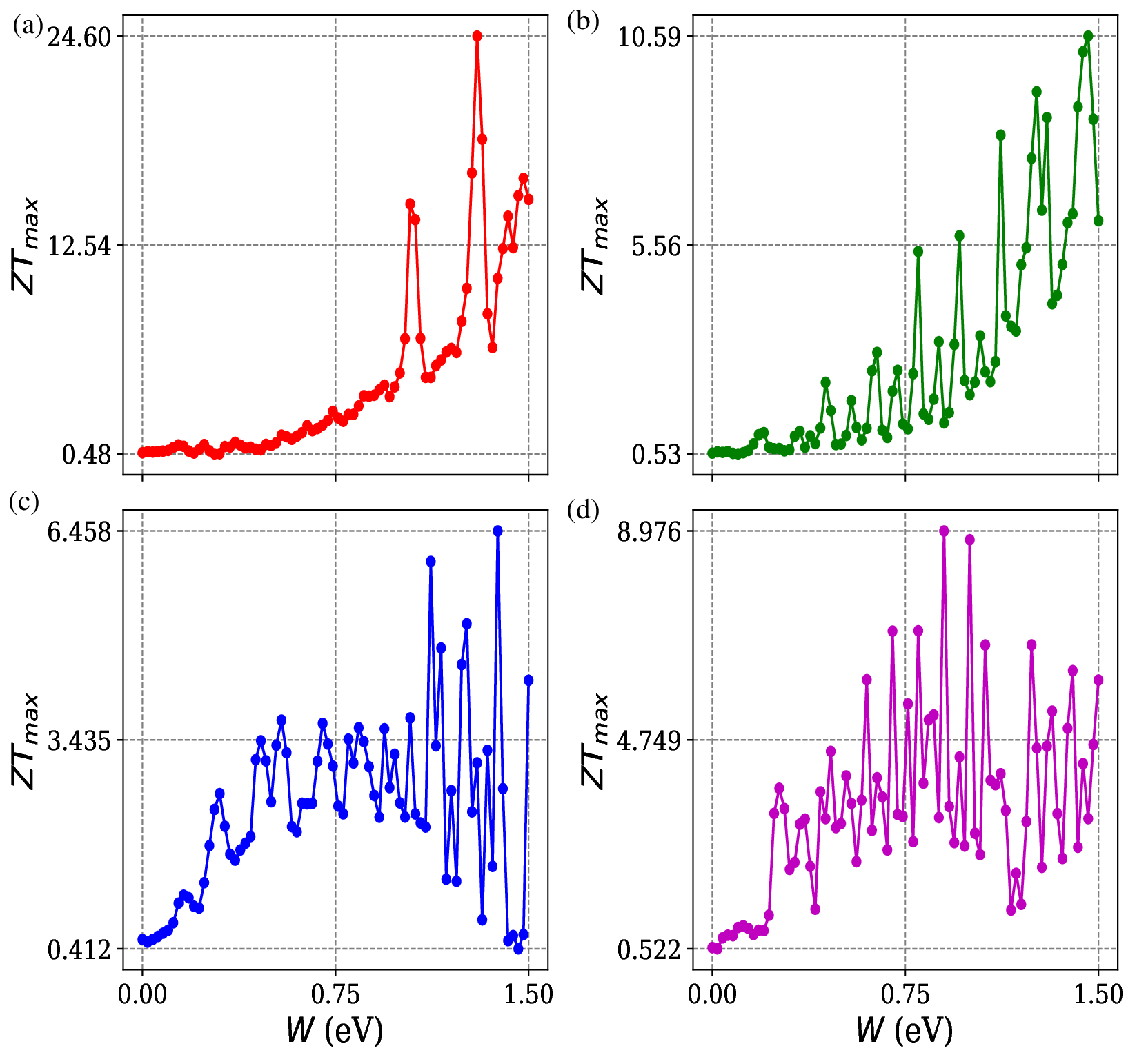}}
\caption{(Color online) Variation of the maximum thermoelectric figure of merit $ZT_{\text{max}}$ as a function of the onsite disorder strength $W$, where $ZT_{\text{max}}$ is computed by scanning over the full range of Fermi energies. Panels (a) and (b) correspond to the non-staggered configuration with $t_1/t_2 > 1$ and $t_1/t_2 < 1$, respectively. Panels (c) and (d) represent the staggered configuration for $t_1/t_2 > 1$ and $t_1/t_2 < 1$, respectively.}
\label{fig11}
\end{figure}

We now turn to the analysis of how the maximum thermoelectric efficiency, characterized by $ZT_{\text{max}}$, varies with the onsite disorder strength $W$. Figure~\ref{fig11} presents a comprehensive comparison for both non-staggered and staggered configurations across different hopping asymmetry regimes.

In the upper panel of Fig.~\ref{fig11}, we examine the non-staggered case. In Fig.~\ref{fig11}(a) For $t_1/t_2 > 1$, $ZT_{\text{max}}$ reaches values as high as 24. This remarkable enhancement can be physically attributed to a nontrivial interplay between disorder-induced localization and the resurgence of delocalized states at moderate disorder strengths.  When the hopping ratio is inverted ($t_1/t_2 < 1$), [Fig.~\ref{fig11}(b)], the system attains a peak $ZT$ value of approximately 10, indicating substantial thermoelectric efficiency. 

Such a nonmonotonic behavior in thermoelectric performance reflects the sensitive dependence of the transmission asymmetry on the disorder profile. In particular, the suppression of thermal conductivity due to localization, coupled with the retention or recovery of electrical conductance through resonant pathways, results in a highly favorable increase in $ZT$. This behavior is particularly evident in the $t_1/t_2 > 1$ non-staggered regime, where a reentrance of delocalized states enhances the asymmetry of the transmission function a key ingredient for high Seebeck response and thus elevated $ZT$.

In the lower panel of Fig.~\ref{fig11}, we depict the $ZT_{\text{max}}$ variation for staggered configurations. For $t_1/t_2 > 1$ [Fig.~\ref{fig11}(c)], the maximum $ZT$ is found to be around 6, whereas for $t_1/t_2 < 1$ [Fig.~\ref{fig11}(d)], it reaches nearly 9. These findings indicate that while the staggered system also exhibits reasonably good thermoelectric behavior, its efficiency is generally lower than its non-staggered counterpart, especially for the $t_1/t_2 > 1$ configuration.

Overall, our analysis suggests that the non-staggered case with $t_1/t_2 > 1$ and the staggered case with $t_1/t_2 < 1$ emerge as the most promising regimes for enhanced thermoelectric performance. The observed trends can be attributed to the subtle balance between disorder-induced localization and the structural asymmetry of hopping amplitudes, which together modulate the electronic and thermal transport characteristics in a nontrivial fashion.

\subsection{Mapping thermoelectric performance across parameter space for non-staggered potentials}

\label{subsec:phase_diagram_nonstaggered}

\begin{figure}[ht]
\centering
\resizebox*{8.5cm}{6.5cm}
{\includegraphics{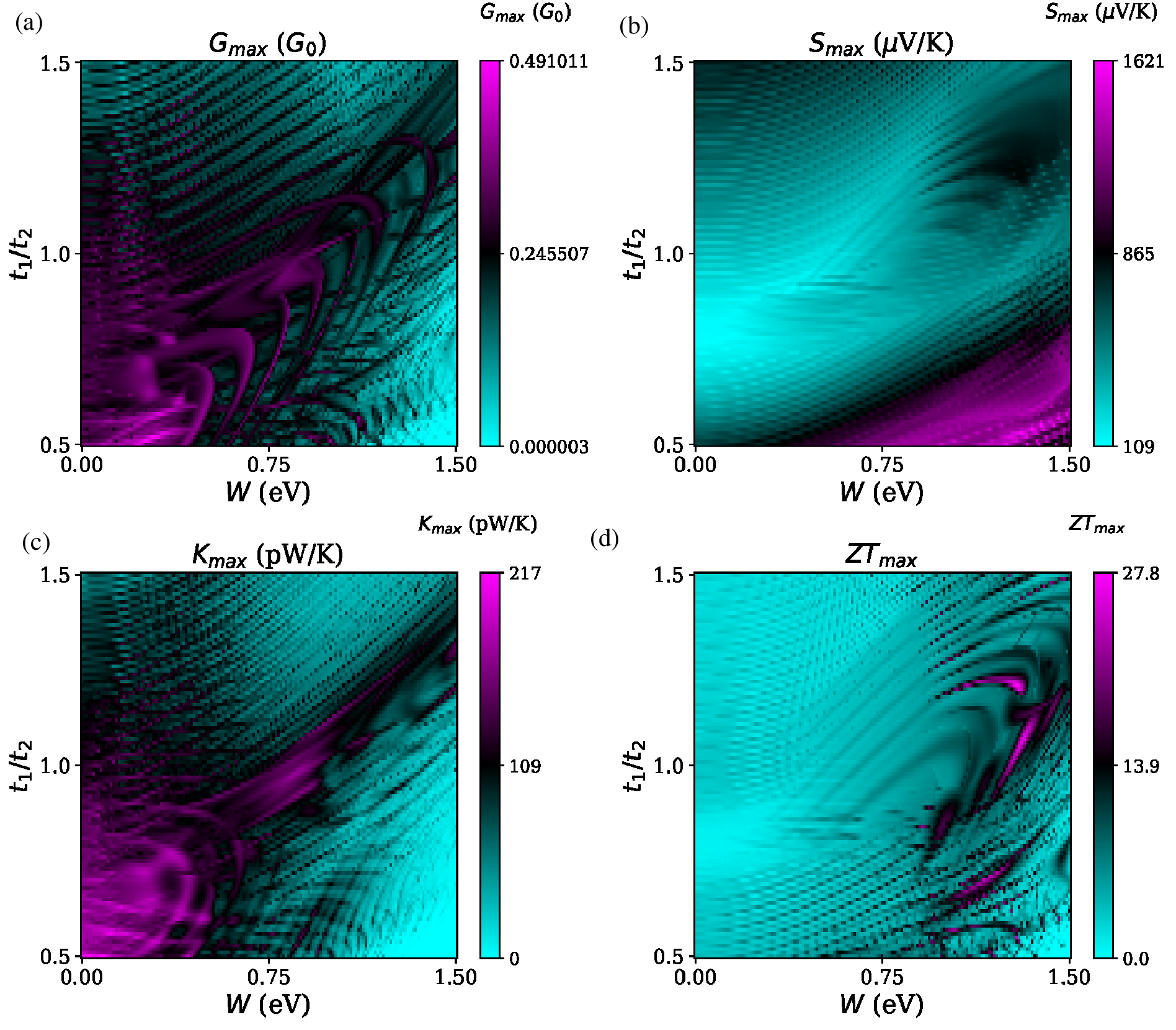}}
\caption{(Color online) Phase diagrams of the maximum values of thermoelectric parameters in the non-staggered configuration. We plot (a) electrical conductance $G_{\text{max}}$, (b) Seebeck coefficient $S_{\text{max}}$, (c) electronic thermal conductance $K^{el}_{\text{max}}$, and (d) figure of merit $ZT_{\text{max}}$ as functions of hopping ratio $t_2/t_1$ and disorder strength $W$.}
\label{fig12}
\end{figure}

To gain a holistic understanding of the thermoelectric behavior across parameter space, we construct phase diagrams for the key thermoelectric quantities in the non-staggered configuration, as shown in Fig.~\ref{fig12}. We fix $t_2 = 0.8$ and allow $t_1$ to vary in the range $[0.5, 1.5]$, while the disorder strength $W$ is simultaneously varied over the same interval.

In Fig.~\ref{fig12}(a), we observe the phase diagram of $G_{\text{max}}$, which exhibits an intriguing reentrant behavior. Specifically, $G_{\text{max}}$ is initially suppressed with increasing $W$, consistent with disorder-induced localization. However, beyond a critical disorder strength ($W \approx 1.3$), it begins to rise again with increasing $t_1$, suggesting the emergence of resonant transmission channels and the partial recovery of metallic behavior a hallmark of the delocalization transition.

Figure~\ref{fig12}(b) shows the variation of the Seebeck coefficient $S_{\text{max}}$, which attains significant values at both ends of the disorder spectrum. High $S$ values are observed for $t_1/t_2 > 1$ at low $W$ and for $t_1/t_2 < 1$ at high $W$, indicating that favorable asymmetry in the energy-dependent transmission function persists in both regimes. Since $S$ is directly related to the spectral asymmetry around the Fermi energy, this result reinforces the notion that the system remains thermoelectrically active over a broad range of parameters.

In Fig.~\ref{fig12}(c), we present $K^{el}_{\text{max}}$, the electronic contribution to thermal conductance. Akin to $G_{\text{max}}$, $K^{el}_{\text{max}}$also displays reentrant characteristics. However, the qualitative behavior diverges in the $t_1/t_2 > 1$ regime, where $K^{el}_{\text{max}}$ remains relatively suppressed despite a rise in $G_{\text{max}}$. This deviation signals a violation of the Wiedemann–Franz (WF) law and hints at the potential for decoupled charge and heat transport an essential requirement for optimizing $ZT$.

Finally, Fig.~\ref{fig12}(d) illustrates the phase diagram of $ZT_{\text{max}}$. The maximum thermoelectric efficiency is achieved near $W \approx 1.3$ for a range of $t_1/t_2$, reaching values around $27.8$. This favorable zone reflects a fine-tuned balance between enhanced Seebeck response, suppressed thermal conductivity, and moderate-to-high electrical conductance.

Collectively, these phase diagrams reveal that the non-staggered configuration supports robust thermoelectric performance across a wide parameter landscape. The coexistence of reentrant conductive behavior and selective suppression of thermal transport is especially conducive to realizing high-efficiency thermoelectric devices in disordered quantum systems.

\subsection{Parameter-space mapping of thermoelectric properties in the staggered potential regime}

\begin{figure}[ht]
\centering
\resizebox*{8.5cm}{6.5cm}
{\includegraphics{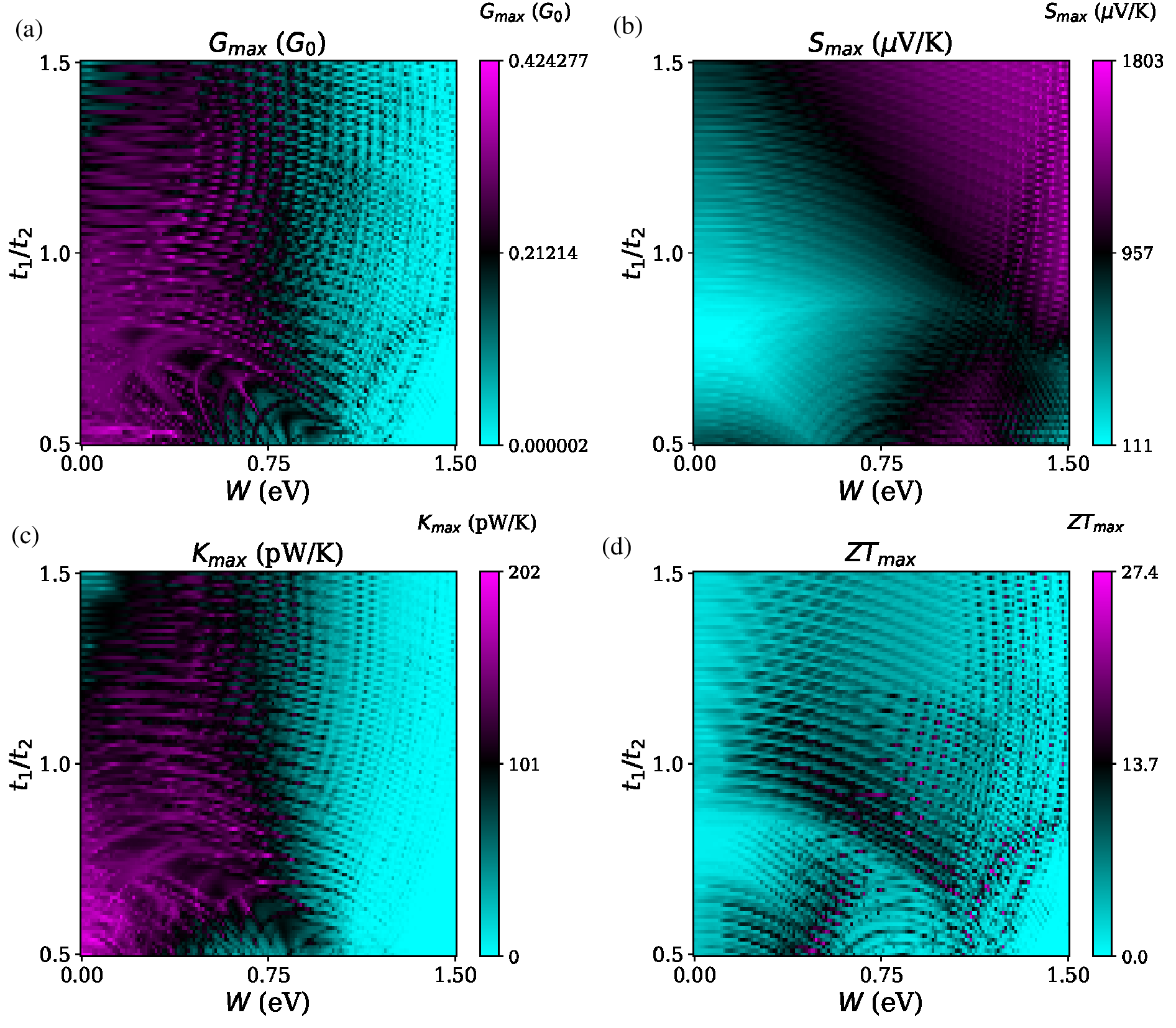}}
\caption{(Color online) Phase diagrams of the maximum thermoelectric parameters in the presence of staggered on-site potentials. The panels show the variation of (a) maximum electrical conductance $G_{\text{max}}$, (b) maximum Seebeck coefficient $S_{\text{max}}$, (c) maximum electronic thermal conductance $K^{el}_{\text{max}}$, and (d) peak thermoelectric efficiency $ZT_{\text{max}}$ as functions of $t_1/t_2$ and disorder strength $W$. The hopping amplitude $t_1$ is fixed at unity.}
\label{fig13}
\end{figure}

In Fig.~\ref{fig13}, we present the phase diagrams of key thermoelectric (TE) quantities, electrical conductance ($G$), Seebeck coefficient ($S$), electronic thermal conductance ($K_{\text{el}}$), and figure of merit ($ZT$) for the case where a staggered modulation of the on-site potential is included. This analysis aims to uncover how the combined effect of quasiperiodic modulation and disorder strength $W$ influences the TE response.

Panel (a) of Fig.~\ref{fig13} displays the variation of $G_{\text{max}}$ with $W$ and the hopping ratio $t_1/t_2$. Interestingly, $G_{\text{max}}$ remains sizable for both low and high disorder values across a broad region of $t_1/t_2$, indicating the reappearance of delocalized electronic states after a localization regime commonly referred to as a reentrant transport behavior. This phenomenon highlights the intricate balance between quantum interference, disorder-induced scattering, and staggered site energy profiles.

Figure~\ref{fig13}(b) shows that $S_{\text{max}}$ also attains elevated values in both weak and strong disorder regimes, particularly when $t_1/t_2 > 1$. The enhanced values of $S$ can be attributed to pronounced asymmetries in the transmission function near the Fermi level, arising from disorder-enhanced energy filtering. Such asymmetry is a known route to optimizing the thermopower in low-dimensional systems.

Panel (c) presents the variation of $K^{el}_{\text{max}}$. Like $G_{\text{max}}$, it exhibits nonmonotonic behavior with increasing $W$, though its suppression at higher $t_1/t_2$ contrasts with the trend observed in $G_{\text{max}}$. This deviation from the Wiedemann–Franz (WF) law suggests the presence of nontrivial transmission pathways, potentially decoupling heat and charge transport channels favorable for achieving higher $ZT$.

Finally, in Fig.~\ref{fig13}(d), we analyze $ZT_{\text{max}}$ across the parameter space. A peak in thermoelectric efficiency is observed around $W \approx 1.3$, where $ZT_{\text{max}}$ reaches values as high as $\sim 27.4$. This peak spans a very few dots of the $t_1/t_2$ axis, reflecting the fact that non-staggared case assures more robust TE performance over a wide range of structural and disorder configurations than the staggared one. The results further establish the role of hopping and onsite modulation in enabling high-efficiency energy conversion in nanostructured systems.

\subsection{Role of contact configuration on the maximization of thermoelectric output}

\begin{figure}[ht]
\centering
\resizebox*{8.5cm}{6.5cm}
{\includegraphics{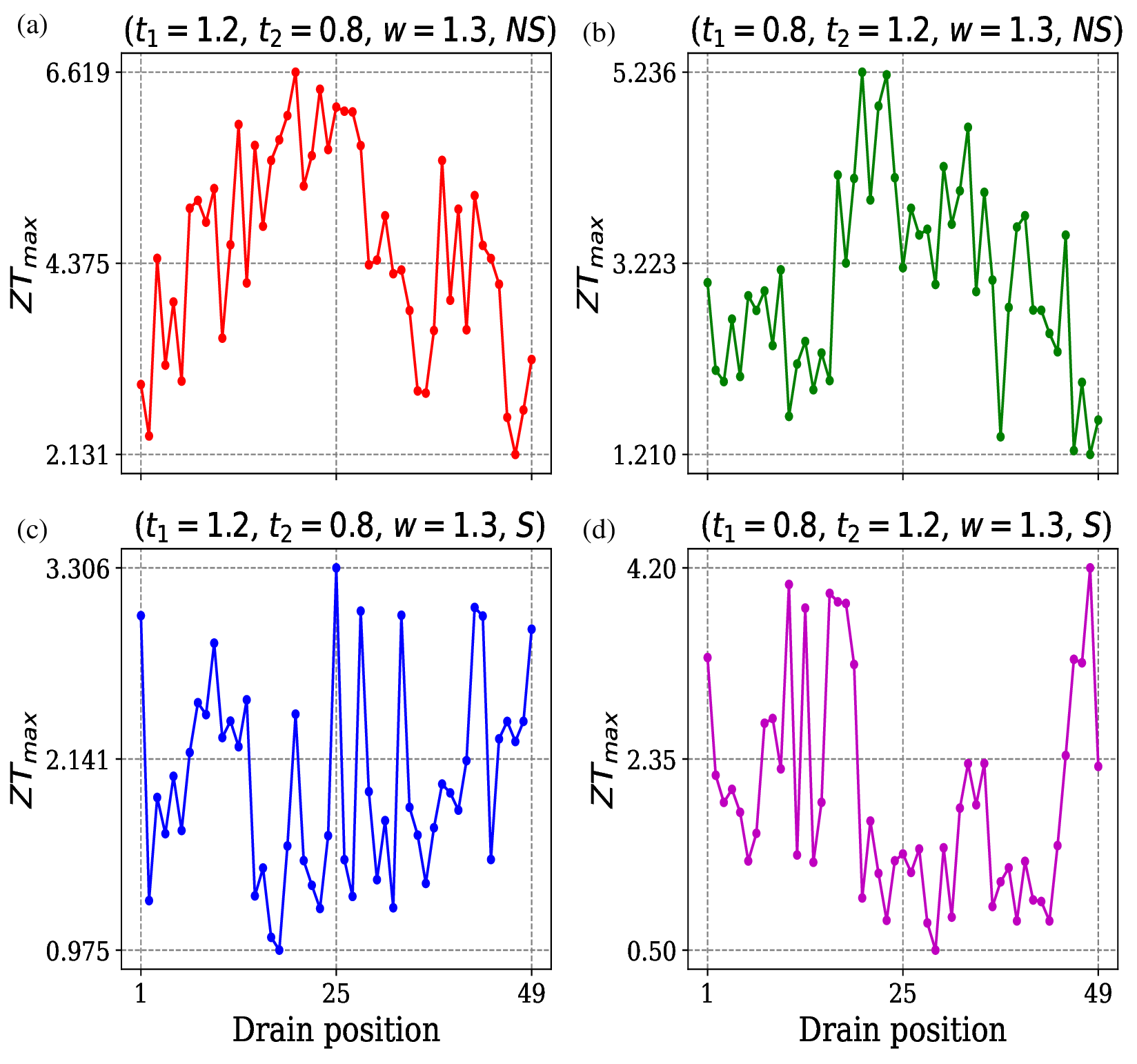}}
\caption{(Color online) Dependence of the peak figure of merit $ZT_{\text{max}}$ on the drain lead position. The upper panels show results for the non-staggered system: (a) $t_1/t_2 > 1$ and (b) $t_1/t_2 < 1$. The lower panels correspond to the staggered case: (c) $t_1/t_2 > 1$ and (d) $t_1/t_2 < 1$.}
\label{fig14}
\end{figure}

To explore the role of quantum interference in our ring-like geometry, we investigate the variation of $ZT_{\text{max}}$ with changes in the drain contact position, as shown in Fig.~\ref{fig14}. The geometry of the system permits multiple interfering electronic paths, and the asymmetry introduced by relocating the drain lead modulates the interference conditions significantly.

Such interference effects are well-documented in mesoscopic systems and are known to influence the line shape of the transmission function near the Fermi energy. In particular, transmission dips or resonances can arise due to constructive or destructive interference, which can in turn boost the Seebeck coefficient and, consequently, $ZT$.

Figure~\ref{fig14}(b) demonstrates that in the non-staggered case with $t_1/t_2 < 1$, repositioning the drain leads to a notable enhancement in $ZT_{\text{max}}$, suggesting that asymmetric configurations are more conducive to energy filtering. This behavior is echoed in the staggered configurations for both $t_1/t_2 > 1$ and $t_1/t_2 < 1$, shown in Figs.~\ref{fig14}(c) and \ref{fig14}(d), respectively. These observations reinforce the idea that structural asymmetry, when combined with quantum interference, can serve as a powerful tuning knob for optimizing thermoelectric performance.

Overall, our results underscore the intricate interplay between disorder, hopping anisotropy, staggered potentials, and quantum interference in modulating thermoelectric behavior. The systematic exploration of these effects reveals parameter regimes that support highly efficient energy conversion, making such systems promising candidates for nanoscale thermoelectric applications. All our essential results are presented in Table~\ref{tab:model_comparison} for non-staggared and staggared cases.
\begin{table*}[ht]
\centering
\renewcommand{\arraystretch}{2}
\resizebox{\textwidth}{!}{%
\begin{tabular}{|c|c|c|c|c|}
\hline
\multicolumn{1}{|c|}{\textbf{Model}} & \multicolumn{2}{|c|}{\textbf{Non-Staggared}} & \multicolumn{2}{|c|}{\textbf{Staggared}}  \\
\hline
\textbf{Hopping configuration} & $t_1 >t_2$ & $t_1<t_2$ & $t_1>t_2$ & $t_1<t_2$ \\
\hline
\textbf{Transmission} & 
\begin{tabular}[c]{@{}l@{}}
$\bullet$Transmission profiles differ significantly \\ spanning almost the entire energy window\\
$\bullet$ Overlapping is minimal
\end{tabular} &
\begin{tabular}[c]{@{}l@{}}
$\bullet$ Transmission profiles largely overlap\\
$\bullet$ Peaks confined to the band edges
\end{tabular} &
\begin{tabular}[c]{@{}l@{}}
$\bullet$ Overlapping transmission peaks\\
$\bullet$ Peaks confined to the band edges
\end{tabular} &
\begin{tabular}[c]{@{}l@{}}
$\bullet$ Broadened and more energetically \\ distributed transmission peaks
\end{tabular} \\
\hline
\textbf{$G_{max}$} &
\begin{tabular}[c]{@{}l@{}}
$\bullet$ Largely dictated by the inherent \\ characteristics of the AAH potential
\end{tabular} &
\begin{tabular}[c]{@{}l@{}}
$\bullet$ Conductance remains lower \\ exhibiting minimal sensitivity \\ to the hopping asymmetry
\end{tabular} &
\begin{tabular}[c]{@{}l@{}}
$\bullet$ Decreases monotonically as \\ disorder increases \\
\end{tabular} &
\begin{tabular}[c]{@{}l@{}}
$\bullet$ Greater resilience to disorder  \\
\end{tabular} \\
\hline
\textbf{$S_{max}$} & 
\begin{tabular}[c]{@{}l@{}}
$\bullet$ Shows decreasing trend with increasing \\ $W$ and gradually reaches to a minimum\\ at $W=0.75$, then rises abruptly \end{tabular}  &
\begin{tabular}[c]{@{}l@{}}
$\bullet$ Gradually rises with increasing $W$ \end{tabular}  &
\begin{tabular}[c]{@{}l@{}}
$\bullet$ Shows increasing trend with \\ increasing  $W$  \end{tabular}  &
\begin{tabular}[c]{@{}l@{}}
$\bullet$ Exhibits a non-monotonic trend \\ with  fall at $W=0.75$ then \\ partial recovery at higher $W$ \end{tabular}  \\
\hline
\textbf{$K_{el_{max}}$} & 
\begin{tabular}[c]{@{}l@{}}
$\bullet$ Exhibits a non-monotonic trend with \\ rising feature then falls around $W=0.6$\\ then rises again afterwards  \end{tabular}  &
\begin{tabular}[c]{@{}l@{}}
$\bullet$ Initially rises and then falls\\  with increasing  $W$  \end{tabular}  &
\begin{tabular}[c]{@{}l@{}}
$\bullet$  Initially rises and then falls \\ with increasing  $W$  \end{tabular}  &
\begin{tabular}[c]{@{}l@{}}
$\bullet$  Initially rises and then falls \\  with increasing  $W$  \end{tabular} \\
\hline
\textbf{$K_{ph_{max}}$} & 
\begin{tabular}[c]{@{}l@{}}
$\bullet$  Falls with increasing  $W$ with \\ a tiny bump around $W=0.4$  \end{tabular} &
\begin{tabular}[c]{@{}l@{}}
$\bullet$  Falls with increasing  $W$   \end{tabular} &
\begin{tabular}[c]{@{}l@{}}
$\bullet$  Falls with increasing  $W$   \end{tabular} &
\begin{tabular}[c]{@{}l@{}}
$\bullet$  Falls with increasing  $W$ with \\ a tiny bump around $W=0.4$  \end{tabular} \\
\hline
\textbf{$ZT_{max}$} & 
\begin{tabular}[c]{@{}l@{}}
$\bullet$  Exhibits a monotonic increase\\ as a function of $W$. \end{tabular} &
\begin{tabular}[c]{@{}l@{}}
$\bullet$  Increases with $W$, accompanied by \\ oscillatory peaks. \end{tabular} &
\begin{tabular}[c]{@{}l@{}}
$\bullet$  Displays a bell-shaped profile with modulated oscillatory\\ features as $W$ increases.\\
$\bullet$ Comparatively lower values than the non-staggered case. \end{tabular} &
\begin{tabular}[c]{@{}l@{}}
$\bullet$  Exhibits a bell-like structure with pronounced\\ oscillatory peaks as $W$ increases.\\ 
$\bullet$The overall magnitude remains lower compared\\ to the non-staggered configuration. \end{tabular} \\
\hline
\end{tabular}
}
\caption{Thermoelectric parameters for non-staggared and staggared cases under diffierent hopping configurations.}
\label{tab:model_comparison}
\end{table*}

\section{Conclusion}
\label{sec:conclusion}

In summary, we have conducted a detailed investigation into the thermoelectric behavior of a Su–Schrieffer–Heeger (SSH) ring structure subject to a modulated on-site potential of the Aubry–André–Harper (AAH) type. Both staggered and non-staggered configurations of the disorder were considered, in conjunction with variable hopping parameters. Through systematic exploration of the parameter space, we have identified regimes where thermoelectric performance, particularly the figure of merit $ZT$, is significantly enhanced.

Our analysis reveals that the competition between quasiperiodic disorder and hopping asymmetry introduces rich transport phenomena, including transitions between conducting and insulating phases. These transitions play a pivotal role in shaping the thermoelectric response, offering a route to optimized energy conversion by selectively tuning structural and energetic parameters.

\vspace{0.2cm}
\noindent\textbf{$\bullet$ Enhancement of $ZT$ via correlated disorder:} The introduction of correlated (AAH-type) disorder modifies the electronic spectrum, introducing asymmetry and energy filtering effects that result in marked improvements in $ZT$. The disorder-induced modulation of the density of states near the Fermi level plays a central role in enhancing thermoelectric efficiency.

\vspace{0.2cm}
\noindent\textbf{$\bullet$ Tunability through hopping anisotropy and disorder:} By varying the ratio $t_2/t_1$ between intercell and intracell hopping amplitudes, we demonstrate that the transport characteristics and thermoelectric parameters can be continuously tuned. This tunability provides a mechanism to engineer favorable conditions for both charge and heat transport.

\vspace{0.2cm}
\noindent\textbf{$\bullet$ Role of quantum interference in transmission control:} The ring geometry inherently supports multiple electronic paths. We explore the effect of drain lead placement, which introduces geometric asymmetry and thereby modifies the interference pattern of electronic wavefunctions. This gives rise to transmission resonances and antiresonances, enhancing $ZT$ via constructive interference and energy filtering.

\vspace{0.2cm}
\noindent\textbf{$\bullet$ Optimized $ZT$ values under specific conditions:} For selective combinations of hopping ratio and disorder strength, we find $ZT$ values exceeding 7, indicating excellent thermoelectric performance. These results emphasize the potential of such systems for practical thermoelectric applications at the nanoscale.

\vspace{0.2cm}
\noindent\textbf{$\bullet$ Implications for thermoelectric design strategies:} Our findings suggest a set of practical design principles for high-efficiency thermoelectric devices. The ability to manipulate quantum interference, localization phenomena, and spectral asymmetry offers a powerful toolkit for developing next-generation energy harvesting technologies.

\vspace{0.2cm}
\noindent\textbf{$\bullet$ Broader impact and future directions:} This work contributes novel insights into the thermoelectric performance of low-dimensional and molecular-scale systems. The demonstrated sensitivity of $ZT$ to structural parameters encourages further exploration of engineered disorder, topology, and multi-terminal configurations to unlock enhanced functionality in nanoscale thermoelectrics.

\noindent Overall, the SSH ring with modulated on-site disorder emerges as a promising prototype system for exploring and harnessing quantum transport effects in thermoelectric applications. Future studies may extend this framework to include phononic effects, electron-phonon interactions, or time-dependent drives to further enhance device performance.

\section*{DATA AVAILABILITY STATEMENT}

The data that underpin the findings of this study can be obtained from the authors upon reasonable request.

\section*{DECLARATION}

\textbf{Conflict of Interest:} The authors affirm that there are no conflicts of interest associated with this work.

\end{document}